\documentclass[fleqn,10pt]{wlscirep}
\usepackage[utf8]{inputenc}
\usepackage[T1]{fontenc}

\usepackage{orcidlink}
\usepackage{tabularx}
\usepackage{booktabs, multirow} % for borders and merged ranges
\usepackage{algorithm}
\usepackage{algpseudocode}
\usepackage{subcaption}
\usepackage{hyperref}

\usepackage{varwidth}
\usepackage{amsmath}
\usetikzlibrary{shapes.geometric, arrows.meta, positioning, calc}

\newcommand{\descbox}[1]{\parbox[t]{0.9\columnwidth}{#1}}

\usepackage[normalem]{ulem}
\usepackage{cancel}
\newif\ifshowrevisions
\showrevisionsfalse
\newcommand{\added}[1]{\ifshowrevisions{\color{blue!75!black}#1}\else#1\fi}
\newcommand{\deleted}[1]{\ifshowrevisions{\color{red!70!black}\ifmmode\bcancel{#1}\else\sout{#1}\fi}\else\fi}
\newcommand{\replaced}[2]{\deleted{#1}\added{#2}}
\title{On the Smart Coordination of Flexibility Scheduling in Multi-Carrier Integrated Energy Systems}

\author[1,*\orcidlink{0009-0002-1905-8221}]{Christian Doh Dinga}
\author[2\orcidlink{0000-0001-6159-041X}]{Sander van Rijn}
\author[3\orcidlink{0000-0002-4014-9294}]{Laurens J. de Vries}
\author[1\orcidlink{0000-0002-5169-7368}]{ Milos Cvetkovic}
\affil[1]{Delft University of Technology, Department of Electrical Sustainable Energy, Delft, 2628 CD, The Netherlands}
\affil[2]{Netherlands eScience Center, Environment and Sustainability Section, Amsterdam, 1098 XH, The Netherlands}
\affil[3]{Delft University of Technology, Department of Engineering Systems and Services, Delft, 2600 GA, The Netherlands}

\affil[*]{corresponding author: c.dohdinga@tudelft.nl}

\begin{abstract}
Coordinating the interactions among flexibility assets in multi-carrier integrated energy systems (MIES) can lead to a cost-efficient energy transition. However, the proliferation of flexibility assets and their growing participation in active demand response increases the complexity of coordinating these interactions. This paper investigates several approaches to model the coordination of flexibility scheduling in MIES with many autonomous flexibility providers. We propose runtime model coupling as an alternative modeling paradigm to overcome the limitations of monolithic centralized co-optimization. Specifically, we introduced two model coupling approaches\textemdash a distributed price-response and a decentralized market auction approach\textemdash to address practical challenges such as preserving the autonomy and privacy of flexibility providers while ensuring scalability. We conduct a quantitative benchmark of these approaches against co-optimization across varying problem sizes, complexities, and computing infrastructures. This benchmark provides new empirical insights into the trade-offs between optimality, autonomy, and scalability that have so far remained unquantified in the energy system literature. We show that model coupling offers a method to balance optimality and realism (autonomy and privacy) while delivering substantial scalability gains. The proposed model coupling approaches are formalized as open-source software with several practical applications: modelers can experiment with different flexibility modeling approaches and choose the one that best matches their modeling objectives and constraints, system scale, and available computing resources; flexibility providers can couple their models to simulate interactions between their systems to make informed operational decisions without disclosing any confidential information.
\end{abstract}

\begin{document}

\flushbottom
\maketitle

\thispagestyle{empty}

\section*{Keywords} \noindent Energy system integration, Multi-energy systems, Distributed energy resources, Demand flexibility, Demand response, Electricity markets, Energy Economics, Multi-agent simulation, Model coupling, Co-simulation, Distributed optimization

% \vspace{-5mm}  %%%%%%%%%%%%%%%%%%%%%%%%%%%%%%%%%%%%%%%%%%%%%%%%%%%%%%%%%%%%%%%%%%%%/
\section*{Nomenclature}

\subsection*{Sets and Indices}
\noindent
\begin{tabbing}
    \hspace{2.5cm} \= \kill % Set tab stop
    $a \in A$ \> \parbox[t]{0.7\columnwidth}{Set of prosumer assets}\\
    $b \in B$ \> \parbox[t]{0.7\columnwidth}{Set of price-quantity bid blocks}\\
    $c \in C$ \> \parbox[t]{0.7\columnwidth}{Set of energy carriers excluding electricity}\\
    $i \in {N}_{g}$ \> \parbox[t]{0.7\columnwidth}{Set of generator agents} \\
    $j \in {N}_{p}$ \> \parbox[t]{0.7\columnwidth}{Set of prosumer agents} \\
    $n \in N$ \> \parbox[t]{0.7\columnwidth}{Set of all agents including generators and prosumers} \\
    $\added{N_{s}^{g} \subseteq N_{g}}$ \> \parbox[t]{0.7\columnwidth}{\added{Set of generator-satellites: generators represented as satellite agents that submit bids}} \\
    $\added{N_{s}^{p} \subseteq N_{p}}$ \> \parbox[t]{0.7\columnwidth}{\added{Set of prosumer-satellites: prosumers represented as satellite agents that submit bids}} \\
    $\added{N_{s} = N_{s}^{g} \cup N_{s}^{p}}$ \> \parbox[t]{0.7\columnwidth}{\added{Set of all satellite agents (both generator and prosumer satellites)}} \\
    $\added{N_{c} = N_{g} \setminus N_{s}^{g}}$ \> \parbox[t]{0.7\columnwidth}{\added{Set of central-model generators: generators whose dispatch is determined directly by the central market-clearing problem}} \\
    $t \in {T}$ \> \parbox[t]{0.7\columnwidth}{Set of time steps} \\
\end{tabbing}
% \vspace{-2mm}  %%%%%%%%%%%%%%%%%%%%%%%%%%%%%%%%%%%%%%%%%%%%%%%%%%%%%%%%%%%%%%%%%%%%/

\subsection*{Parameters}
\noindent
\begin{tabbing}
    \hspace{2.5cm} \= \kill % Set tab stop
    $\alpha_{i}, \beta_{i}$ \> \descbox{Generator quadratic and linear cost coefficients \quad [$\text{€}/MW^2h$, $\text{€}/MWh$]} \\
    $\lambda^{forecast}_{j, t}$ \> \descbox{Electricity price forecast \quad [$\text{€}/MWh$]} \\
    $\pi_{j, b, t}$ \> \descbox{Price of bid block $b$ \quad [$\text{€}/MWh$]} \\
    $\mu_{c, t}$ \> \descbox{Cost of energy carrier \quad [$\text{€}/MWh$]} \\
    $\replaced{\delta^{p}}{\delta^{p}_{a}}$ \> \descbox{Binary parameter: equals 1 if the conversion process \added{at asset $a$} involves electricity, 0 otherwise \quad [-]} \\
    $\gamma^{q}$ \> \descbox{Binary parameter: equals 1 if the conversion \\ process involves heat, 0 otherwise \quad [-]} \\
    $P^{D}_{j, t}$ \> \descbox{Base demand of electricity \quad [$MW$]} \\
    $R^{D}_{c, j, t}$ \> \descbox{Base demand of non-electric carrier \quad [$MW$]} \\
    $\replaced{n^{E}_{a, j}}{\eta^{E}_{a, j}}$ \> \descbox{Electrical efficiency of prosumer asset} \\
    $\replaced{n^{r}_{a, j}}{\eta^{r}_{a, j}}$ \> \descbox{Non-electric efficiency of prosumer asset} \\
    $\replaced{n^{r, ch}, n^{r, dc}}{\eta^{r, ch}, \eta^{r, dc}}$ \> \descbox{Storage charge/discharge efficiency} \\
    $\overline{I_{a, j}} $ \> \descbox{Maximum capacity of converter asset \quad [$MW$]} \\
    $\overline{S^{r}_{j}} $ \> \descbox{Maximum capacity of storage \quad [$MW$]} \\
    $\underline{G_{i,t}}, \overline{G_{i,t}}$ \> \descbox{Generator minimum, maximum available capacity \quad [$MW$]} \\
    $\underline{E^{r,S}_{j}}, \overline{E^{r,S}_{j}}$ \> \descbox{Storage minimum, maximum energy capacity \quad [$MWh$]} \\
    $\underline{E^{r,F}_{j}}, \overline{E^{r,F}_{j}}$ \> \descbox{Flexible demand min/max energy \quad [$MWh$]} \\
    $\rho$ \> \descbox{Penalty parameter in price-response coupling} \\
    $k$ \> \parbox[t]{0.7\columnwidth}{Iteration number in price-response coupling} \\
    % $\deleted{I^{k-1}}$ \> \descbox{\deleted{Imbalance at iteration $k-1$ \quad [$MWh$]}} \\
    % $\deleted{D^{k-1}}$ \> \descbox{\deleted{Dual residual at iteration $k-1$ \quad [$MWh$]}} \\
    $\added{\Delta t}$ \> \descbox{\added{Time step length \quad [$h$]}} \\
    $\added{\overline{\Phi}_{j, b, t}}$ \> \descbox{\added{Maximum quantity submitted for prosumer bid block $b$ \quad [$MW$]}} \\
\end{tabbing}

\vspace{-2mm}  %%%%%%%%%%%%%%%%%%%%%%%%%%%%%%%%%%%%%%%%%%%%%%%%%%%%%%%%%%%%%%%%%%%%
\subsection*{Primal and dual variables}
\noindent
\begin{tabbing}
    \hspace{2.5cm} \= \kill % Set tab stop
    $\lambda_t$ \> \descbox{Electricity (market clearing) price \quad [$\text{€}/MWh$]} \\
    $g_{i,t}$ \> \descbox{Generator dispatch \quad [$MW$]} \\
    $p_{j,t}$ \> \descbox{Prosumer net electric power \quad [$MW$]} \\
    $p^{gen}_{c, a, j, t}$ \> \descbox{Prosumer electricity generated locally \quad [$MW$]} \\
    $p^{con}_{a, j, t}$ \> \descbox{Electricity consumed from the grid \quad [$MW$]} \\
    $p^{ch}_{a, j, t}, p^{dc}_{a, j, t}$ \> \descbox{ (EV) battery charge/discharge \quad [$MW$]} \\
    $q^{st}_{j, t}$ \> \descbox{Heat generation from solar thermal \quad [$MW$]} \\
    $r^{ch}_{a, j, t}, r^{dc}_{a, j, t}$ \> \descbox{Non-electric storage charge/discharge \quad [$MW$]} \\
    $p^{F}_{j, t}, r^{F}_{j, t}$ \> \descbox{Flexible demand of energy carrier \quad [$MW$]} \\
    $x_{c, a, j, t}$ \> \descbox{Consumption rate of energy carrier \quad [$MW$]} \\
    $r_{a, j, t}$ \> \descbox{Generation rate of energy carrier \quad [$MW$]} \\
    $e^{r,F}_{j, t}$ \> \descbox{Energy level of flexible demand \quad [$MWh$]} \\
    $e^{r,S}_{j, t}$ \> \descbox{Energy level of (battery) storage \quad [$MWh$]} \\
    $\phi_{j, b, t}$ \> \descbox{Power quantity of bid block $b$ \quad [$MW$]} \\
    % ==============================================================================
    % \\ % Space to avoid overlap with copyright
    % \\
    % ==============================================================================
\end{tabbing}
% \vspace{-10mm}  %%%%%%%%%%%%%%%%%%%%%%%%%%%%%%%%%%%%%%%%%%%%%%%%%%%%%%%%%%%%%%%%%%%%

\section*{Introduction}

The growing electrification of energy demand and proliferation of distributed energy resources (e.g., heat pumps, electric vehicles, electrolyzers, etc.) increases the interactions and interdependencies between energy sectors and carriers \cite{MANCARELLA_2014_EGY, Martinez_2020, garg_2025_nsr}, resulting in multi-carrier integrated energy systems (MIES). While energy system integration offers new sources of flexibility, the inclusion of many different types of distributed energy resources (herein, referred to as flexibility resources) increases the complexity of the energy system \cite{OCONNELL_2014_RSER, PONCELET_2020_AE}. This is because it becomes challenging to coordinate interactions among demand-side flexibility resources \cite{li_2025_optimization}. Unlike traditional, centralized supply-side flexibility resources such as conventional generators that are (partially) controlled by the system operator, demand-side flexibility resources are controlled by prosumers who have full autonomy over their systems \cite{Mohsenian_2010_TSG, GHATIKAR_2016_autonomous}. This implies that prosumers can directly, or through aggregators, participate in active demand response where they schedule their power consumption and/or generation predominantly based on price signals \cite{HAQUE_SEGN_2017, Bruninx_2018_TSE}. As shown in Harder \textit{et al.} \cite{Harder_2023_EI}, excluding the autonomy of prosumers in energy system models can result in lower operational profits, sending misleading investment signals to flexibility providers, which are expected to play a crucial role in the energy transition \cite{LUND_2015_review_flex, irena_2018_flexibility, Strbac_2020_IOP}. Moreover, disregarding the autonomy of prosumers fails to capture the potential consequences of demand response. For example, recent studies have shown that uncoordinated response from a large pool of demand-side flexibility resources can aggravate congestion in the electricity grid if prosumers simultaneously ramp up consumption in response to low electricity prices \cite{Dinga_ISGT_2024, STAWSKA_2021_EP}. Therefore, energy system integration brings along a methodological challenge\textemdash the challenge of coordinating the interactions among \replaced{an arbitrarily large number of}{many} different types of flexibility resources (owned and operated by prosumers with full autonomy) to optimally schedule flexibility.

In the literature, the conventional approach for modeling the coordination of flexibility scheduling (also referred to as demand response) in MIES is based on combined optimization (co-optimization). In this approach, all subsystems (prosumers or simply agents) are represented within a single, monolithic computational tool, and their interactions are coordinated by optimizing a global utility function, typically system-wide costs. Co-optimization has been widely applied in various MIES contexts. For instance, it has been used to coordinate flexibility scheduling in regional systems consisting of interconnected electricity, heat, and natural gas networks \cite{WANG_2019_AE, ZHANG_2024_SETA}. It has also been extended to account for uncertainty in the operation of MIES \cite{Martinez_2019_TSG, Li_2022_TIA}, as well as to assess the flexibility potential arising from increased electrification of heat demand in MIES \cite{PATTEEUW_2015_AE, Zhang_2019_TSG}. Numerous other studies \cite{Zhang_2016_TPS, Gottwalt_2017_TSG, DONG_2023_AE} adopt similar monolithic co-optimization approaches to model flexibility in MIES.

Although widely used, co-optimization has several drawbacks. First, its top-down control strategy assumes that agents give up control over their system, and that local control (dispatch) decisions are made centrally. This assumption takes away the autonomy of prosumers, which, as we already argued, is a key characteristic that reflects the operational realities of flexibility providers. Therefore, while co-optimization could be a valid approach for modeling a single agent's flexibility scheduling problem, it is, however, not suitable for studying flexibility in a MIES where many agents with full autonomy interact. Second, co-optimization scales poorly with increasing problem size (e.g., number of agents), since all dispatch decisions are computed centrally (see Figure~\ref{Co_optimization_interactions}). As the number of agents and the complexity of their models grow, the size of the resulting optimization problem increases significantly, which may lead to higher computational burden in terms of memory requirements and solution time. This challenge can be further exacerbated when long time horizons or intertemporal constraints\textemdash which are required to accurately represent flexibility\textemdash are considered. While advances in solver technology and high-performance computing (HPC) can alleviate some of these issues, it remains unclear to what extent co-optimization remains computationally tractable for large-scale MIES with many heterogeneous agents. Finally, perhaps, the most practical limitation of co-optimization is the need for information sharing, including confidential data. This is because to centrally compute dispatch decisions, co-optimization requires complete information of all agents (see Figure~\ref{Co_optimization_interactions}), including the techno-economic parameters and internal states of their assets. This raises a privacy issue, as in reality, agents will not be willing to share such private information about their system for confidentiality reasons.

To address the aforementioned limitations of monolithic co-optimization in modeling the coordination of flexibility scheduling in large-scale MIES with many autonomous agents, we propose to use model coupling. Model coupling is a modeling paradigm that simulates an integrated system by combining independent, standalone models through the exchange of interface variables enabled by software techniques and algorithms \cite{gusain_2022_dissertation, Palensky_2017_IEM}. \added{The literature on model coupling spans a spectrum of strategies that differ fundamentally in when information is exchanged among coupled models. At one end of this spectrum are offline model coupling strategies, commonly known as soft-linking \cite{Deane_2012_EGY, Frigg_2022_ae, ZHANG_2024_soft_linking, RAJAKOVIC_2025_ecmx, ROSENDAL_2025_soft_linking}. In offline or soft-linking coupling strategies, each model is executed in isolation without any interaction during runtime, and thus information exchange occurs only after each model completes its run; consequently, coupling is reduced to post-hoc data exchange. At the middle of this spectrum, are runtime model coupling strategies, also referred to as co-simulation \cite{Palensky_2017_IEM, Palensky_2024_PEM}. Here, coupled models exchange interface variables during execution, while model interaction and time progression are managed by an explicit algorithmic coordination mechanism \cite{Palensky_2017_IEM, Palensky_2024_PEM}. As a result, the computations (decisions) of each coupled model become mutually dependent and are updated dynamically throughout the simulation run. At the other far end of this spectrum are hard-linking strategies that integrate all submodels into a single monolithic model, eliminating inter-model interfaces entirely, thereby inheriting the properties and limitations of centralized co-optimization discussed above.} \added{Among these model coupling strategies, runtime model coupling is particularly well-suited for modeling flexibility in MIES with many autonomous agents. This is because offline coupling strategies cannot capture the dynamic, mutually dependent interactions that arise in flexibility scheduling, where the interactions among many models (agents) must be coordinated at runtime. Accordingly, in the remainder of this paper, we use the term model coupling to refer specifically to runtime coupling (co-simulation).}

More broadly, our motivation for proposing model coupling for studying flexibility in MIES with many autonomous agents is threefold: \replaced{autonomy, privacy, and scalability.}{\added{autonomy (the locus of decision-making, that is, whether agents' dispatch decisions are computed locally or centrally, and whose utility that decision optimizes), privacy (the volume and granularity of techno-economic information that an agent must disclose to a coordinator or to other agents for the coordination to function), and scalability (how computational requirements, such as runtime and memory usage, grow with increasing problem size and complexity.)}} First, regarding autonomy, model coupling can \replaced{preserve}{enhance the} autonomy \added{of flexibility providers compared to monolithic co-optimization}. This is because model coupling allows each agent to retain its own model and thus its decision-making-related computations, while interactions among agents are coordinated through information exchange protocols. As a result, agents can locally optimize their utility and compute their dispatch decisions, thereby enabling a more realistic representation of prosumer behavior and their interactions. Second, concerning privacy, because coupled models exchange only interface variables, private information such as techno-economic parameters or system internal states remains local to each model (agent), \replaced{eliminating the need for prosumers to disclose any confidential information about their systems}{reducing the volume and granularity of information that prosumers must disclose regarding their systems}. Finally, regarding scalability, the modularity of model coupling allows each agent (subsystem) to be simulated in its native tool using dedicated solvers; from a computational point of view, this modularity enables parallelism as independent models (agents) can be simulated concurrently, thereby promising substantial gains in computational efficiency and scalability.

\added{
We note that the autonomy and privacy advantages of model coupling are relative rather than absolute, and are best understood with respect to a baseline. Throughout this paper, the baseline is monolithic co-optimization. Accordingly, the assessments reported in Table~\ref{tab:table1} should be interpreted as qualitative comparisons of how the three coordination approaches differ in the degree of autonomy retained by agents and the amount of information that must be disclosed to achieve coordination. A rigorous quantitative assessment of either concept is beyond the scope of the present work. For example, autonomy could be quantified through measures of the deviation between locally preferred and centrally imposed dispatch decisions, while privacy could be quantified using information-theoretic metrics of disclosure \cite{Han_2017_TAC}. Developing and evaluating such metrics remains an important direction for future research. Moreover, increased privacy does not imply complete protection against the inference of information. As documented in optimization and control literature~\cite{Han_2017_TAC, Aswani_2018_OR, Dvorkin_2020_TPS}, the repeated exchange of dispatch and price signals in the price-response approach may, in principle, allow a coordinator to infer agent cost parameters through inverse-optimization techniques. Similarly, detailed price-quantity bids in the market auction approach may reveal information about an agent's underlying cost structure to a sufficiently informed market operator. Developing stronger guarantees against such inference attacks, for example, through differential privacy or related privacy-preserving mechanisms, also lies beyond the scope of this work and constitutes another promising direction for future research.
}

Model coupling has recently gained increasing attention in the energy systems literature as a means of enabling high-fidelity energy system assessments~\cite{WIDL2022_segn, lund_2023_model_coupling_energy, ROSENDAL_2025_soft_linking}. However, existing research on coupling energy models has focused primarily on the development of coupling infrastructures rather than on the coordination logic governing interactions between coupled models~\cite{mosaik_2019, GUSAIN_EnergySym, HELICS_2024}. Consequently, the question of how to coordinate the interactions among operational models of different prosumers at runtime remains largely unexplored and poorly understood. This study, therefore, addresses this gap by investigating alternative strategies for coupling operational energy models of different types of flexible prosumers at runtime to effectively capture the dynamic, mutually dependent interactions required for coordinating flexibility scheduling in MIES. Depending on the required information exchange among the coupled models and on how their interactions are managed during runtime (i.e., the coordination mechanism), fundamentally different coordination architectures emerge, most notably, distributed and decentralized architectures. In this work, we introduce two model coupling approaches, each corresponding to one of the coordination architectures: a price-response approach based on a distributed coordination architecture, and a market auction approach based on a decentralized coordination architecture. \added{Our architecture naming follows the convention in the energy system optimization literature \cite{Han_2017_TAC, Kargarian_2018_distributed_vs_decentralized} in which distributed requires a coordinator agent that broadcasts a signal (penalty or incentive) to steer the system towards equilibrium, and decentralized requires no such coordinator agent. It is worth noting that other alternative coordination paradigms exist for studying multi-agent energy systems, notably multi-agent reinforcement learning \cite{Vazquez_Canteli_2019_AE} and game-theoretic equilibrium models \cite{Saad_2012_SPM}. However, both lie outside the scope of this paper as our primary focus is on coordination paradigms based on model coupling.}

% =================================== price-response ==================
In the proposed price-response model coupling approach (see Figure~\ref{price_response_interactions}), agents locally optimize their dispatch decisions and interact to cooperatively optimize a global utility. The required information exchange includes: (i) point-wise demand-supply time series computed locally by each agent as a response to price time series broadcast by a coordinator agent; and (ii) the system imbalance computed as the difference between total demand and supply. Given that agents' dispatch must satisfy the system-wide power balance constraint that couples their decisions, the broadcast prices and agents' dispatch decisions are iteratively updated based on the imbalance until a consensus price is reached at which the imbalance is below a specified threshold. Generally, provided an equilibrium exists, this iterative procedure converges to a solution (almost) equal to that of centralized monolithic co-optimization. However, although optimal, this approach also has some drawbacks. First, the autonomy of agents is somewhat partial since their dispatch decisions are penalized over iterations to align them towards cooperatively optimizing a global utility. Second, the iterative nature of the coordination mechanism can lead to high computational costs, since decisions are recomputed over the full simulation horizon at each iteration. As a result, the approach can scale poorly if applied to large-scale MIES with long time horizons. Moreover, convergence is not guaranteed in general, particularly when agents' optimization problems are non-convex.

% =================================== market auction ==================
The market auction approach\textemdash which uses a decentralized coordination architecture instead  (see Figure~\ref{market_auction_interactions})\textemdash presents an alternative model coupling strategy that can overcome the limitations of both monolithic co-optimization and the price-response approach. The market auction approach is characterized by the following: (i) it allows for full autonomy in agents' local dispatch decisions in a more realistic way. This is because it relaxes the assumption of a central agent optimizing dispatch decisions in a top-down control manner, or a coordinator agent penalizing agents' local dispatch decisions over iterations to align them towards a global optimum. This, however, comes with a sacrifice in optimality compared to co-optimization and the price-response approach; (ii) the coordination mechanism here simulates market auctions by matching agents' bids through social welfare maximization, requiring that information is exchanged within the resolution and horizon of the market clearing. This promises significant gains in scalability since the full horizon is partitioned into smaller windows and simulated in a rolling manner; (iii) it only requires agents to share their demand and supply bid functions that internalize their techno-economic and intertemporal constraints. How they construct these bids is their confidential trade strategy encapsulated in their respective models. In doing so, agents encode all their system details in a bid function. This \replaced{preserves}{enhances their privacy (relative to co-optimization), as it avoids the need to share any further information other than bids}; (iv) interestingly, exchanging bid functions rather than point-wise time series (as in the price-response approach) eliminates the need for iterations, which can be prone to non-convergence, especially in the presence of non-convexities.

% % ============================= Summary of the three approaches =============================
The fundamental differences between the coordination approaches we have discussed so far are summarized in Table~\ref{tab:table1}. As shown, these approaches differ mainly in their modeling paradigm (monolithic vs. model coupling), coordination architecture, coordination mechanism, required information exchange, and their representation of autonomy in agents' local decision-making.

% =================== Table on taxonomy of coordination architectures =================== % 
\begin{table}[htbp]
    \caption{Taxonomy of various approaches for modeling the coordination of flexibility scheduling in MIES}
    \label{tab:table1}
    \centering
    \footnotesize
    \renewcommand{\arraystretch}{1.2}
    
    % \begin{tabularx}{\textwidth}{lXXX}
    \begin{tabularx}{\textwidth}{p{0.18\textwidth}XXX}
        \toprule
        \multirow{2}{*}{\textbf{Criteria}} & \multicolumn{3}{c}{\textbf{Coordination approach}} \\
        \cmidrule(lr){2-4}
        & \textbf{Co-optimization} 
        & \textbf{Price-response} 
        & \textbf{Market-auction} \\
        \midrule
        Modeling paradigm 
        & Monolithic model
        & Model coupling
        & Model coupling \\
    
        Coordination architecture 
        & Centralized coordination 
        & Distributed coordination 
        & Decentralized coordination \\

        Coordination mechanism 
        & Centrally optimize a global utility through a one-shot optimization 
        & Cooperatively optimize a global utility through iterations 
        & Match agents' bids through rolling horizon-based market clearing \\

        Required information exchange 
        & Complete information of agents, including confidential parameters 
        & Interface variables: dispatch and price time series, and system imbalance 
        & Interface variables: demand and supply bid functions, and market price \\
        
        Autonomy in agent's local decision-making 
        & No autonomy: top-down control signals from a central optimizer 
        & Partial autonomy: local optimization based on consensus price updates 
        & Full autonomy: local optimization based on expected market price \\
        
        \bottomrule
    \end{tabularx}
\end{table}
% ========================= end table ========================= %

% ============================= Contributions =============================
In sum, the main contributions of this paper with respect to the scientific literature include:

\begin{enumerate}
    \item We propose a model coupling paradigm for modeling the coordination of flexibility scheduling in MIES. Unlike traditional monolithic approaches, model coupling (i) allows for a more realistic representation of prosumer behavior since it retains their autonomy (ii) \replaced{preserves privacy as coupled models only share interface variables}{enhances privacy since coupled models only exchange interface variables}, and (iii) shows promising improvement in scalability since each coupled model can be simulated in parallel in its native tool. We introduce two such model coupling approaches with fundamentally different coordination architectures: a distributed price-response approach and a decentralized market-auction approach. 

    \item We conduct a systematic benchmarking of monolithic co-optimization and the two proposed model coupling approaches across varying problem sizes (spanning 30–350 agents and simulation horizons from one day to one week), problem complexity (LP vs MILP), and computing infrastructures (laptop vs HPC) . This benchmarking provides new empirical evidence on the computational trade-offs between autonomy (accuracy), theoretical optimality, and scalability across different flexibility modeling approaches that have so far remained unquantified in the energy system modeling literature.

    \item We formalize and implement our proposed model coupling approaches as new open-source computational frameworks \cite{doh_dinga_2025_price_response, Dinga_2025_coupling} that are freely available for the scientific community to explore and experiment with different approaches for modeling flexibility scheduling in MIES. The code for running all experiments conducted in this paper is also made open source and archived on Zenodo \cite{Dinga_2025_smart_coordination} to ensure that the results presented here are reproducible by third parties.

    \item From a practical perspective, this work provides guidelines for modeling flexibility in MIES. More specifically, it demonstrates how different modeling architectures can be used to balance realism (in terms of agents' autonomy and privacy), theoretical optimality, and computational tractability, thereby providing insights into which approach is most suitable depending on modeling objectives, system scale, and available computational resources.

\end{enumerate}

% ======================== Description of the paper structure ========================
The remainder of this paper is structured as follows. The methodology section presents a detailed description and mathematical formulations of the three coordination approaches for modeling flexibility scheduling in MIES. The experiment design section describes the experimental setup, case study, and data sources. The results and discussion section presents the findings. Finally, the conclusion section summarizes the key insights from this study.

\section*{Methodology}
\begin{figure}[htbp]
    \centerline{\includegraphics[width=0.65\textwidth]{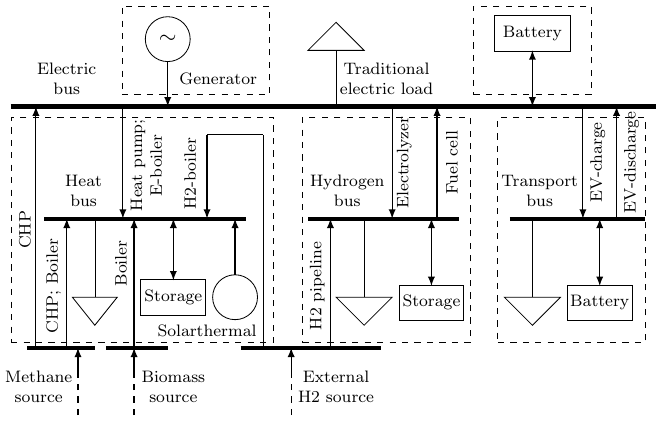}}
    \caption{Schematic of the multi-carrier integrated energy system (MIES) used as the reference for mathematical formulations throughout this paper. The electricity, heat, hydrogen, and transport subsystems are coupled through Power-to-X and X-to-Power conversion technologies. Each subsystem enclosed in a dashed rectangle represents an autonomous agent.}
    \label{MIES_fig}
\end{figure}

This section provides a description of the physical and economic interactions between flexibility assets and agents, and the problem formulations under the three coordination approaches. While any coordination objective could be used, we assume that the objective of each agent is to minimize costs or maximize revenues. The formulations presented here are restricted to LP and MILP problems for the sake of simplicity, and are based on the MIES shown in Figure~\ref{MIES_fig}, where each subsystem (enclosed in dashed rectangles) representing an agent is treated as a copper-plate following Brown \textit{et al.} \cite{Brown_2018_ENGY}. \added{We further assume that all prosumers and generators are connected to a single electric bus. This modeling choice primarily aims to isolate the coordination layer, which is the central object of study in this work. Specifically, we aim to compare the scaling behavior of the three coordination paradigms under a common physical representation of the MIES. This abstraction is also consistent with the current European electricity market design, in which day-ahead market clearing is performed at the zonal resolution, while intra-zonal congestion is managed ex-post through redispatch \cite{Poplavskaya_2020_AE}. Nevertheless, the formulations presented here can be easily extended to spatially-resolved MIES; see, for example, Neumann \textit{et al.} \cite{NEUMANN_2023_Joule}. The implications of relaxing the copper-plate assumption, and the extent to which the findings of this study transfer to spatially-resolved MIES, are discussed in detail in the \textit{Results and Discussion} section.}

% Introducing spatially-resolved network constraints would create locational dependencies and nodal price heterogeneity that interact differently with each coordination architecture, thereby obscuring a controlled comparison of optimality and scalability across the three approaches.

The MIES shown in Figure~\ref{MIES_fig} consists of heat, hydrogen, and transport systems coupled to the electricity system through Power-to-X and X-to-Power technologies. The electricity system has generators, batteries, and traditional loads that represent demand from passive consumers. Each physical system enclosed in a dashed rectangle represents an agent with full autonomy. Taking the heat subsystem as an example, the heat prosumer has a heat load that can be satisfied through multiple assets. For example, when electricity prices are lowest, the heat prosumer could choose to dispatch electric heating assets (e.g., heat pumps), shifting demand from other carriers to electricity to minimize costs. However, when electricity prices are highest, the prosumer can, for instance, dispatch its methane or biomass-based CHP asset to meet heat demand while also generating revenues from electricity sales. The heat prosumer also has a heat storage to buffer heat production and increase flexibility by decoupling demand and supply in time.

\subsection*{Co-optimization}
\begin{figure}[htbp]
    \centerline{\includegraphics[width=0.7\textwidth]{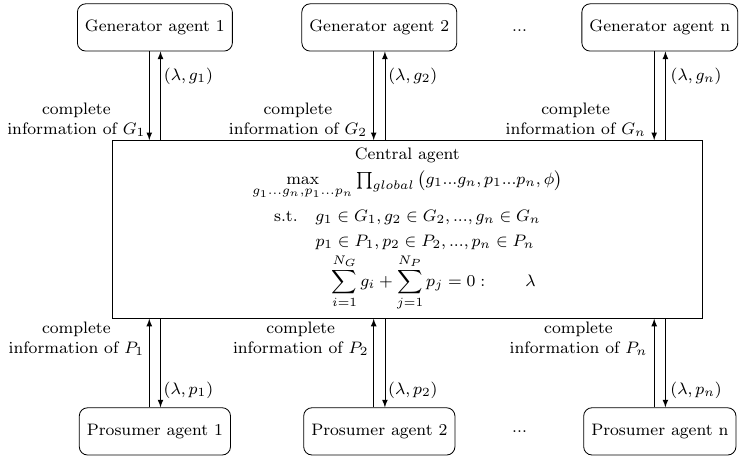}}
    % height=0.3\textheight
    \caption{Interactions under centralized monolithic co-optimization. All generator and prosumer agents disclose their complete techno-economic parameters and internal states to a central agent, who solves a single system-wide optimization problem. The resulting per-agent optimal dispatch is then sent to each agent to be executed in a top-down control manner.}
    \label{Co_optimization_interactions}
\end{figure}

Under co-optimization, the coordination problem is formalized as a centralized optimization problem where the dispatch of all assets is determined by minimizing the operational costs to satisfy the demand for all energy carriers. The coordination architecture that illustrates how the interactions in co-optimization would happen in real-world is depicted in Figure~\ref{Co_optimization_interactions}.

First, generator and prosumer agents share all their local information (e.g., techno-economic parameters, internal states of their systems, loads, load shifting potential, etc.) with a central agent. The central agent then computes all dispatch decisions centrally and sends the optimal operational schedules to all agents to be executed in a top-down manner. Mathematically, the problem can be written as follows.

\allowdisplaybreaks
\begin{flalign}
    & {\text{min}} \quad \sum_{i, t} \Bigl(\alpha_{i} \cdot g^{2}_{i,t} + \beta_{i} \cdot g_{i,t}\Bigr) + \sum_{c, a , j, t} \Bigl(\mu_{c, t} \cdot x_{c, a, j, t}\Bigr)  && \label{co_optim_obj_func} \\
    & \text{subject to:} \nonumber \quad \\
    & \sum^{N_{G}}_{i=1} g_{i,t} + \sum^{N_{P}}_{j=1} p_{j, t} = 0 \qquad \qquad \forall t:  \quad \lambda_t && \label{system_electric_power_balance_cons} \\
    % =====================================================================
    & \left.
        \begin{aligned}
            & p_{j,t} = \sum_{c, a} p^{gen}_{c, a, j, t} \added{+ \sum_{a} p^{dc}_{a, j, t}} - \replaced{\Bigl(\sum_{a} p^{con}_{a, j, t} +  p^{ch}_{a, j, t} \Bigr)}{\sum_{a} \Bigl(p^{con}_{a, j, t} +  p^{ch}_{a, j, t} \Bigr)}
            -  \Bigl(p^{F}_{j, t} + P^{D}_{j, t} \Bigr) \qquad \qquad \forall j, t \hfill \label{prosumer_electric_power}
        \end{aligned} \quad
    \right. \\
    % =====================================================================
    & p^{gen}_{c, a, j, t} = \replaced{n^{E}_{a, j}}{\eta^{E}_{a, j}} \cdot x_{c, a, j, t} \deleted{+  p^{dc}_{a, j, t}} \qquad \qquad \forall c, a , j, t  && \label{conversion_from_other_carriers_to_elec_cons} \\
    % =====================================================================
    & \left.
        \begin{aligned}
            & r_{a, j, t} = \replaced{n^{r}_{a, j}}{\eta^{r}_{a, j}} \cdot \Bigl( \sum_{c\neq r} x_{c, a, j, t} + \replaced{\delta^{p}}{\delta^{p}_{a}} \cdot p^{con}_{a, j, t}\Bigr)  \qquad \qquad \forall r \in \{q,h\}, a, j, t \hfill \label{energy_conversion_cons}
        \end{aligned} \quad
    \right. \\
    % =====================================================================
    % =====================================================================
     & \left.
        \begin{aligned}
            & \sum_{a} r_{a, j, t} + \gamma^{q} \cdot q^{st}_{j, t} + r^{dc}_{j, t} - r^{ch}_{j, t} = r^{F}_{j, t} + R^{D}_{c=r, j, t} \qquad \qquad \forall r \in\{q,h\}, j, t \hfill \label{prosumer_local_q_and_h_resource_balance}
        \end{aligned} \quad
     \right. \\
    % =====================================================================
    & e^{r,F}_{j, t} = e^{r,F}_{j, t-1} + r^{F}_{j, t} \added{\cdot \Delta t} \qquad \qquad \forall  r \in\{p,q,h\}, j, t && \label{flex_demand_store_energy_dynamics_cons} \\
    & e^{r,S}_{j, t} = e^{r,S}_{j, t-1} + \replaced{n^{r,ch}}{\eta^{r,ch}} \cdot r^{ch}_{j, t} \added{\cdot \Delta t} - \frac{\replaced{r^{ds}_{j, t}}{r^{dc}_{j, t}} \added{\cdot \Delta t}}{\replaced{n^{r,dc}}{\eta^{r,dc}}} \qquad \qquad \forall  r \in\{p,q,h\}, j, t && \label{storage_energy_dynamics_cons} \\
    & \underline{G_{i,t}} \leq g_{i,t} \leq \overline{G_{i,t}} \qquad \qquad \forall i, t && \label{gen_dispatch_limit_cons} \\
    & 0 \leq x_{c, a, j, t} \replaced{,}{} \quad p^{con}_{a, j, t} \leq \overline{I_{a, j}}  \qquad \qquad \forall c, a, j, t  && \label{converter_dispatch_limit_cons} \\
    & 0 \leq r^{ch}_{j, t} \replaced{,}{} \quad r^{dc}_{j, t} \leq \overline{S^{r}_{j}}  \qquad \qquad \forall r \in\{p,q,h\},j, t && \label{storage_dispatch_limit} \\
    & \underline{E^{r,S}_{j}} \leq e^{r,S}_{j, t} \replaced{,}{} \leq \overline{E^{r,S}_{j}}  \qquad \qquad\forall r \in\{p,q,h\},j, t  && \label{storage_SOC} \\
    & \underline{E^{r,F}_{j}} \leq e^{r,F}_{j, t} \replaced{,}{} \leq \overline{E^{r,F}_{j}}  \qquad \qquad \forall r \in\{p,q,h\},j, t  && \label{flexible_demand_feasible_region}
\end{flalign}

The objective function defined by Equation~\ref{co_optim_obj_func} minimizes the cost of electricity generation and the consumption costs of externally priced energy carriers, such as methane and biomass. The end-use energy carriers: electricity, heat, and hydrogen, are denoted by $p,q,h$, respectively. Equation~\ref{system_electric_power_balance_cons} ensures that electricity generation and consumption must be balanced at all time steps. The prosumer electric power defined in Equation~\ref{prosumer_electric_power} is the difference between the electricity generated locally and that consumed from the grid or market; it is negative for net consumption and positive for net generation. Constraints defined by Equation~\ref{conversion_from_other_carriers_to_elec_cons} and Equation~\ref{energy_conversion_cons} capture the conversions between energy carriers; the local balance of heat and hydrogen is enforced by Equation~\ref{prosumer_local_q_and_h_resource_balance}; Equation~\ref{flex_demand_store_energy_dynamics_cons} and Equation~\ref{storage_energy_dynamics_cons} capture the energy dynamics of flexible demand and storage; Equations~\ref{gen_dispatch_limit_cons} to \ref{storage_dispatch_limit} limit the dispatch of generators and prosumer assets within their technical capacities; and finally, the constraints defined by Equation~\ref{storage_SOC} and Equation~\ref{flexible_demand_feasible_region} model the minimum and maximum state-of-charge of storage and the feasible energy levels of flexible demands. We assume that flexible demand can be shifted in time but is not curtailable. \added{We exclude load shedding from the scope of \emph{flexibility} considered in this paper, as it represents a reduction in service rather than a temporal redistribution of energy use. Otherwise, modeling such behavior would require additional assumptions about willingness-to-pay or value-of-lost-load that are outside the scope of this work.} \added{The non-electric carrier (methane, biomass, and hydrogen) prices $\mu_{c, t}$ are deterministic, time-varying exogenous parameters and thus, are not formed endogenously by market clearing in the model. Only the electricity price $\lambda_{t}$\textemdash the dual variable of the electric power balance in Equation~\ref{system_electric_power_balance_cons}, and thus the shadow price of electricity\textemdash is determined endogenously in the model.} \added{Furthermore, the objective in Equation~\ref{co_optim_obj_func} does not include a utility term for flexible demand, that is, load shifting is assumed to be free of inconvenience cost. Generators are also not subject to ramping constraints and thus can adjust intertemporal dispatch instantaneously. While these assumptions are consistent with our objective of isolating the effects of different coordination approaches from additional modeling complexities, such assumptions will tend to favor co-optimization; therefore, any optimality gap reported between co-optimization and the two proposed model coupling approaches should be interpreted as a lower bound.}

\subsection*{Price-response model coupling}
\begin{figure*}[htbp]
    \centering
    \captionsetup{justification=centering}
    \includegraphics[width=0.92\textwidth]{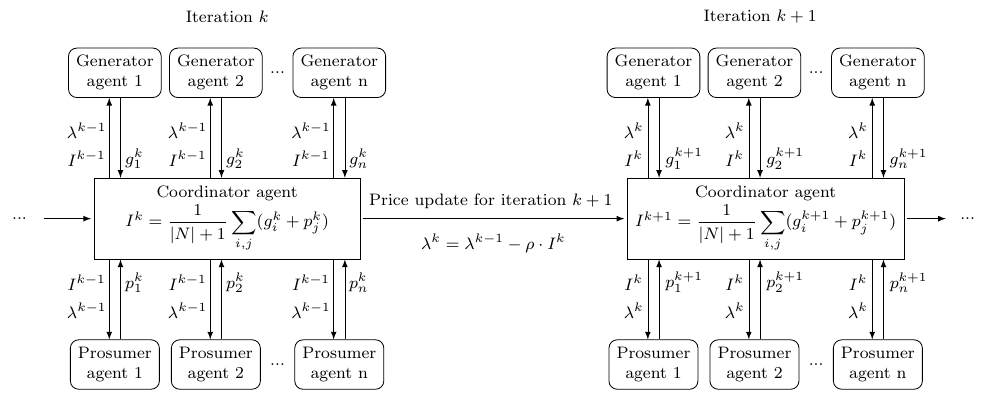}
    % height=0.25\textheight
    \caption{Interactions under price-response model coupling. Each generator and prosumer agent solves its own local optimization problem and exchanges only its dispatch decision (techno-economic parameters and internal states remain local) and a consensus price signal with the coordinator agent, who iteratively updates this price signal until convergence.}
    \label{price_response_interactions}
\end{figure*}

In the price-response approach, the coordination problem is formalized as a set of optimization problems, each representing the dispatch optimization of a generator or prosumer. A visualization of the interactions in this approach is illustrated in Figure~\ref{price_response_interactions}. As shown, it is an iterative process that starts with a coordinator agent broadcasting a price time series and system imbalance to agents. Given this information, agents run their local optimization to determine their optimal dispatch. Next, they report back their electric power generation or consumption time series to the coordinator agent, who then calculates the new system imbalance. If the stopping criteria (system imbalance defined by Equation~\ref{admm_system_imbalance} and change in agents' decisions over iterations defined by Equation~\ref{dual_residual}) is not below the predefined threshold, a new price timeseries is computed and broadcast, and the process repeats again. \added{The pseudo-code of the price-response coordination mechanism is shown in Algorithm~\ref{price_response_admm}}. In what follows, we define the optimization problem of each agent type in iteration $k$.

\begin{algorithm}
    \caption{Price-response coordination mechanism}
    \label{price_response_admm}
    \algrenewcommand\algorithmicindent{1.0em}
    \begin{algorithmic}[1]
        \Require Initial price $\lambda^{0}_{t}$ for all $t$; initial penalty $\rho^{0} > 0$; maximum iteration count $K_{\max}$; convergence tolerance $\epsilon$; initial dispatch $g^{0}_{i,t} = 0$ for all $i \in N_{g}$ and $p^{0}_{j,t} = 0$ for all $j \in N_{p}$; initial imbalance $I^{0}_{t} = 0$ for all $t$
        \State $k \gets 0$
        \While {$k < K_{\max}$}
            \State $k \gets k + 1$
            \Statex \hspace{-0.6em}\textit{Step~1: Local agent updates (executed in parallel).}
            \ForAll {$i \in N_{g}$ \textbf{in parallel}}
                \State Compute $g^{k}_{i,t}$ by solving Equation~\ref{generator_agents_objective_function_admm}
            \EndFor
            \ForAll {$j \in N_{p}$ \textbf{in parallel}}
                \State Compute $p^{k}_{j,t}$ by solving Equation~\ref{prosumer_agents_objective_function_admm}
            \EndFor
            \Statex \hspace{-0.6em}\textit{Step~2: Coordinator gathers responses and computes the system imbalance.}
            \State $I^{k}_{t} \gets \dfrac{1}{|N|+1} \cdot \Bigl(\sum_{i \in N_{g}} g^{k}_{i,t} + \sum_{j \in N_{p}} p^{k}_{j,t}\Bigr) \quad \forall t$ \Comment{Eq.~\ref{admm_system_imbalance}}
            \Statex \hspace{-0.6em}\textit{Step~3: Primal stopping criterion.}
            \State $\Psi^{k} \gets \big\|I^{k}\big\|_{2}$ \Comment{Eq.~\ref{primal_residual}}
            \Statex \hspace{-0.6em}\textit{Step~4: Dual stopping criterion (well-defined for $k \geq 2$).}
            \If {$k \geq 2$}
                \State Compute $\widetilde{\Psi}^{k}$ from Equation~\ref{dual_residual}
            \Else
                \State $\widetilde{\Psi}^{k} \gets +\infty$
            \EndIf
            \Statex \hspace{-0.6em}\textit{Step~5: Price update and broadcast to agents.}
            \State $\lambda^{k}_{t} \gets \lambda^{k-1}_{t} - \rho^{k-1} \cdot I^{k}_{t} \quad \forall t$ \Comment{Eq.~\ref{admm_price_update}}
            \Statex \hspace{-0.6em}\textit{Step~6: Dynamic penalty-parameter update (residual balancing).}
            \If {$\Psi^{k} > 10 \cdot \widetilde{\Psi}^{k}$}
                \State $\rho^{k} \gets 2 \cdot \rho^{k-1}$ \Comment{Primal stagnant: amplify the penalty}
            \ElsIf {$\widetilde{\Psi}^{k} > 10 \cdot \Psi^{k}$}
                \State $\rho^{k} \gets \rho^{k-1} / 2$ \Comment{Dual stagnant: relax the penalty}
            \Else
                \State $\rho^{k} \gets \rho^{k-1}$ \Comment{Residuals balanced: keep the penalty}
            \EndIf
            \Statex \hspace{-0.6em}\textit{Step~7: Joint convergence check.}
            \If {$\Psi^{k} \leq \epsilon$ \textbf{and} $\widetilde{\Psi}^{k} \leq \epsilon$}
                \State \textbf{break}
            \EndIf
        \EndWhile
        \State \Return committed dispatch $\{g^{k}_{i,t}, p^{k}_{j,t}\}_{i,j,t}$, equilibrium prices $\{\lambda^{k}_{t}\}_{t}$, iteration count $k$, and residual history $\{\Psi^{k}, \widetilde{\Psi}^{k}\}_{k=1}^{K}$
    \end{algorithmic}
\end{algorithm}

\subsubsection*{Generator agents}
Each generator agent maximizes its revenues from electricity generation within its technical limits by solving the following optimization problem.

\begin{flalign}
    & \left.
    \begin{aligned}
        & {\text{max}} \sum_{t} \Bigl(\lambda^{k-1}_t - (\alpha_{i} \cdot g^{k}_{i,t} + \beta_{i})\Bigr) \cdot g^{k}_{i,t} 
        \replaced{+}{-} \frac{\rho}{2} \cdot \Big\| g^{k}_{i,t} - \Bigl( g^{k-1}_{i,t} -\replaced{I^{k-1}}{I^{k-1}_t}\Bigr)\Big\|^{2}_{2} \hfill \label{generator_agents_objective_function_admm} \\
        & \text{subject to:}  \quad  \quad \text{constraints defined by} \quad\text{Equation}~\ref{gen_dispatch_limit_cons}
    \end{aligned}
    \right\} \quad \forall i \in {N}_{g}
\end{flalign}

The first term in the objective represents revenue maximization, while the second term is an augmented penalty term. The penalty serves to dampen oscillations by constraining abrupt changes in agents' local dispatch decisions across iterations. This helps drive the iterative process towards an equilibrium state by penalizing agents' dispatch decisions based on the system imbalance. This enforces agents to align their local decisions towards a global optimum while still allowing them to optimize locally, resulting in partial autonomy in their decision-making. \added{Although the penalty is applied across all time steps, it acts only as an iterative proximal regularizer that vanishes at equilibrium; it is zero at equilibrium. Consequently, the penalty term does not affect the dispatch of agents at the equilibrium solution; see Boyd~\textit{et al.}~\cite{Boyd_2011} for more details on this algorithm.}

\subsubsection*{Prosumer agents}
Each prosumer agent maximizes revenues from electricity generated locally, or minimizes the consumption costs of energy carriers, subject to its technical constraints. The following problem is established for each prosumer.

\begin{flalign}
    & \left.
    \begin{aligned}
        & {\text{min}} \replaced{\sum_{c, a, t} - \lambda^{k-1}_t \cdot p^{k}_{j,t} + (\mu_{c, t} \cdot x_{c, a, j, t})}{\sum_{t} \Bigl( -\lambda^{k-1}_t \cdot p^{k}_{j,t} \Bigr) + \sum_{c, a, t} \Bigl( \mu_{c, t} \cdot x_{c, a, j, t} \Bigr)} 
        + \frac{\rho}{2} \cdot \Big\| p^{k}_{j,t} - \Bigl( p^{k-1}_{j,t} - \replaced{I^{k-1}}{I^{k-1}_t}\Bigr)\Big\|^{2}_{2}  \label{prosumer_agents_objective_function_admm} & \\
        & \text{subject to:}  \quad  \text{constraints defined by} \quad \text{Equations}~\ref{prosumer_electric_power} - \ref{storage_energy_dynamics_cons} \quad \text{and} \quad \text{Equations}~\ref{converter_dispatch_limit_cons} - \ref{flexible_demand_feasible_region}
    \end{aligned}
    \right\} \quad \forall j \in {N}_{p}
\end{flalign}

As mentioned earlier, when net consuming, $p^{k}_{j,t}$ is negative, and the term related to electricity in the objective becomes cost minimization; when net generating, $p^{k}_{j,t}$ is positive, and the electricity-related term becomes revenue maximization.

\subsubsection*{The coordinator agent}
The role of the coordinator is to calculate the imbalance of the system in each iteration as in Equation~\ref{admm_system_imbalance}, and to update prices for the next iteration according to Equation~\ref{admm_price_update}.
\begin{flalign}
    & \replaced{I^{k}}{I^{k}_t} = \frac{1}{\left|{N}\right| + 1} \cdot \Bigl(\sum_{\replaced{i, t}{i}} {g^{k}_{i, t}} + \sum_{\replaced{j, t}{j}} {p^{k}_{j, t}}\Bigr) \added{\quad \forall t} \label{admm_system_imbalance} \\
    &  \lambda^{k+1}_{t} = \lambda^{k}_{t} - \rho \cdot \replaced{I^{k}}{I^{k}_t} \label{admm_price_update}
\end{flalign}

The price update is done in the direction that reduces the imbalance: for example, if there is excess demand in iteration $k$, the price in iteration $k+1$ is increased, and vice versa. \added{The imbalance is normalized by the number of subproblems contributing primal updates: $|N| + 1$ represents generator and prosumer agents plus the coordinator agent that performs the price update.}

\added{The iterative process is controlled by two stopping criteria: a primal stopping criterion $\Psi^{k}$ and a dual stopping criterion $\widetilde{\Psi}^{k}$. The primal residual, defined per time step, measures the violation of the power balance constraint. The primal residual stopping criteria at iteration $k$ is therefore defined as the $\ell_2$-norm of the imbalance residual vector $I^{k} = (I^{k}_{1}, \ldots, I^{k}_{|T|})$ as in Equation~\ref{primal_residual}.}

\begin{flalign}
    \added{\Psi^{k} \;=\; \big\| I^{k} \big\|_{2} \;=\; \sqrt{\sum_{t} \bigl(I^{k}_{t}\bigr)^{2}}.} \label{primal_residual}
\end{flalign}

\added{The dual residual, defined per agent and per time step, measures the change in agents' decisions between consecutive iterations. It is multiplied by the penalty factor $\rho$ to link the change in decisions to the change in prices. Similarly, the dual residual stopping criteria is defined as the $\ell_2$-norm of the dual residual as in Equation~\ref{dual_residual}.}

\begin{flalign}
    \begin{aligned}
    \added{\widetilde{\Psi}^{k} \;=\;}
    & \added{\Big\| \rho \cdot \bigl[(g^{k}_{i, t} - I^{k}_{t}) - (g^{k-1}_{i, t} - I^{k-1}_{t}) \bigr] \Big\|_{2}}
    \quad +\, & \added{\Big\| \rho \cdot \bigl[(p^{k}_{j, t} - I^{k}_{t}) - (p^{k-1}_{j, t} - I^{k-1}_{t}) \bigr] \Big\|_{2}.}
    \end{aligned} \label{dual_residual}
\end{flalign}

\added{The algorithm terminates when $\Psi^{k} \leq \epsilon$ and $\widetilde{\Psi}^{k} \leq \epsilon$ are simultaneously satisfied. In the standard ADMM formulation presented by Boyd~\textit{et al.}~\cite{Boyd_2011}, separate tolerances are typically used for the primal and dual residuals, since they are not dimensionally equivalent and may differ in scale. Here, however, we adopt a single tolerance threshold $\epsilon$ for both $\Psi^{k}$ and $\widetilde{\Psi}^{k}$ as a practical convergence criterion. This is made possible by using a dynamic penalty update rule, which dynamically adjusts $\rho$ to keep the primal and dual residuals at comparable magnitudes during the iterative process. The dynamic penalty update strategy is often used to accelerate convergence speed \cite{Bruninx_2020_EE, Hoschle_2018_TPS}.}

 % Empirically, we found that this adaptive residual-balancing strategy also improved convergence behavior. The dynamic penalty parameter update rule we adopted is described as follows.

\added{
\subsubsection*{The dynamic penalty parameter update rule}
The penalty parameter $\rho$ in Equations~\ref{generator_agents_objective_function_admm} and~\ref{prosumer_agents_objective_function_admm} plays a critical role in the convergence speed of the iterative procedure. A value of $\rho$ that is too small leaves agents' local dispatch decisions only weakly coupled to the coordinator's price signal (slow primal convergence), whereas a value of $\rho$ that is too large over-penalizes deviations and forces premature consensus before the coordinator has had a chance to refine the price (slow dual convergence). The classical residual-balancing strategy of Boyd \textit{et al.}~\cite{Boyd_2011} addresses this by adjusting $\rho$ between iterations in response to the relative magnitudes of the primal and dual residuals (see Step 6 of Algorithm~\ref{price_response_admm}): doubling $\rho$ whenever the primal stopping criterion $\Psi^{k}$ exceeds ten times the dual stopping criterion $\widetilde{\Psi}^{k}$, halving $\rho$ whenever $\widetilde{\Psi}^{k}$ exceeds ten times $\Psi^{k}$, and otherwise keeping $\rho$ unchanged. We adopt this rule in Step~6 of Algorithm~\ref{price_response_admm}.

In our empirical experiments, the dynamic rule materially increases the fraction of convex problem instances that converge within the iteration limit and reduces the mean iteration count when compared to a static-$\rho$. We emphasize that improving the underlying ADMM algorithm is not the focus of this paper: we adopt the dynamic-$\rho$ update rule as a well-known modification that improves convergence \cite{Bruninx_2020_EE, Hoschle_2018_TPS}, and otherwise treat the algorithm as a fixed building block of the price-response coordination architecture. Readers interested in a deeper treatment of penalty parameter selection and convergence acceleration techniques for ADMM are referred to Boyd \textit{et al.}~\cite{Boyd_2011} and the surrounding literature.
}

% Our use of model coupling here is largely agnostic to the specific consensus-ADMM variant chosen at the coordinator agent; any reasonable rule could, in principle, be substituted.

\subsection*{The price-response model coupling software}
We implement the price-response model coupling approach as a new open-source Python-based software package called Demoses-admm \cite{doh_dinga_2025_price_response}. Demoses-admm provides a streamlined API for the exchange of messages\textemdash formatted as one-dimensional arrays\textemdash among the coupled models (agents). \added{Inter-model communication is fully synchronous: within every iteration, the coordinator blocks until every agent has returned its updated dispatch before updating and proceeding to the next iteration. The software leverages the consensus form of ADMM, with a residual-balancing rule for the penalty parameter $\rho$, which is a well-known modification of the standard ADMM algorithm \cite{Boyd_2011}}. The implementation is structured in a modular manner, allowing individual models to be defined independently while remaining fully compatible with the communication protocol. This enables the flexible integration of an arbitrary number of heterogeneous models, as well as parallel execution of agent optimizations within each iteration. Demoses-admm abstracts key functionalities such as price update mechanisms, penalty parameterization, iteration control, message passing, and convergence monitoring. This design makes the framework highly customizable, allowing users to tune coordination settings while focusing on higher-level tasks, such as the formulation of agent-specific models, rather than the underlying coordination logic and algorithmic overhead.

% \added{Inter-agent communication in Demoses-admm is fully synchronous: within every iteration of Algorithm~\ref{price_response_admm}, the coordinator broadcasts the current price $\lambda^{k}_{t}$ and imbalance $I^{k-1}_{t}$ to all agents, then blocks until \emph{every} agent has returned its updated dispatch before computing the next imbalance and price update. The coordinator never proceeds with stale or partial data; this synchronization is essential to the convergence guarantees of consensus-ADMM and reflects our experimental setup, in which we deliberately rule out asynchronous-update variants to keep the comparison with co-optimization on equal footing.}

\subsection*{Market auction model coupling}
\begin{figure}[htbp]
    \centering
    \captionsetup{justification=centering}
    \includegraphics[width=0.6\textwidth]{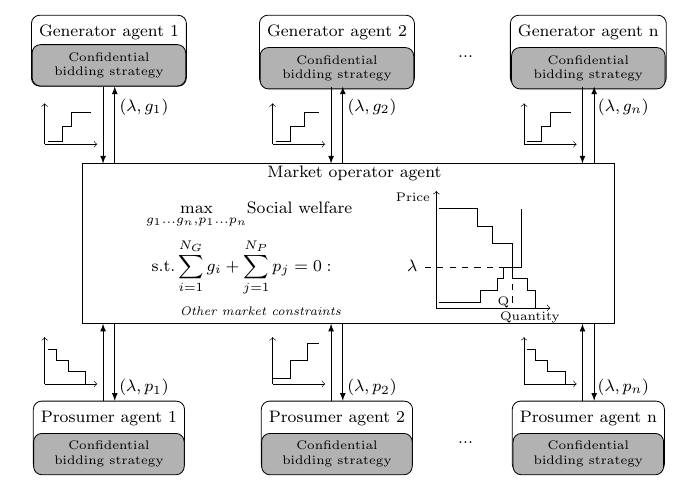}
    % height=0.35\textheight
    \caption{Interactions under market auction model coupling. Each generator and prosumer agent constructs its demand/supply bids based on a local price forecast and submits them to the market operator; techno-economic parameters and internal states remain local. The market operator then clears the market and returns the clearing price and accepted quantities to each agent.}
    \label{market_auction_interactions}
\end{figure}

In the market auction approach, agents interact through a decentralized market mechanism, where dispatch and prices are determined through auctions as illustrated in Figure~\ref{market_auction_interactions}. First, agents run their local optimization based on electricity price forecast scenarios to determine their demand and or supply bids. These bids, which reflect their willingness to generate or consume electricity at various price levels, are then submitted to the market operator agent. Next, the market operator clears the market, and the resulting market-clearing prices and dispatch quantities are communicated back to agents. In this setup, we refer to the market operator model as the central model and the agents’ models as satellite models. The central market operator model and the agents’ satellite models are described below, including their mathematical formulations.

\subsubsection*{The central model}
\replaced{The central model is an electricity market-clearing model that matches agents' demand and supply bids. Since the focus of our study is on demand-side flexibility, we assume a simple bidding strategy for generators: generators always bid their available capacity at true marginal costs. Hence, in the market auction formulation, generator constraints defined by Equation~\ref{gen_dispatch_limit_cons} are directly included in the central model. It is worth noting that, if the bidding behavior of specific generators is of interest, such generators can instead be modeled as satellite agents with full autonomy, analogous to how prosumer agents are represented here. Under the above assumptions, the mathematical formulation of the central model is given as follows.}{The central model is an electricity market-clearing model that matches agents' demand and supply bids with the goal of maximizing social welfare. Within the framework, generators and prosumers can be represented as satellites that submit their own bids to the central model, or alternatively folded directly into the central market-clearing problem. Let $N_{s} = N_{s}^{g} \cup N_{s}^{p}$ denote the set of satellite agents, partitioned into generator-satellites $N_{s}^{g} \subseteq N_{g}$ (those whose bids are computed by their local optimization as in Equation~\ref{generator_satellite_obj}) and prosumer-satellites $N_{s}^{p} \subseteq N_{p}$. Let $N_{c} = N_{g} \setminus N_{s}^{g}$ denote the set of generators whose dispatch is determined directly inside the central market-clearing problem; for these generators, their constraints can be directly included in the central model. The specific partition between $N_{c}$ and $N_{s}^{g}$ used in our experiments is reported in the \textit{Experiment Design} section. Under this generic formulation, the central market-clearing problem can be written mathematically as follows.}

\begin{flalign}
    & {\text{max}} \quad \sum_{\replaced{j, b, t}{j \in N_s, b, t}} \Bigl(\pi_{j, b, t} \cdot \phi_{j, b, t}\Bigr) - \sum_{\replaced{i, t}{i \in N_c, t}} \Bigl(\alpha_{i} \cdot g^{2}_{i,t} + \beta_{i} \cdot g_{i,t}\Bigr)  \hfill \label{max_sw_obj} & \\
    & \text{subject to:} \qquad \qquad \text{constraints defined by} \quad\text{Equation}~\ref{gen_dispatch_limit_cons}\added{\text{ for } i \in N_c} \nonumber \quad \\
    & \replaced{\sum^{N_{G}}_{i=1}}{\sum_{i \in N_c}} g_{i,t} + \replaced{\sum^{N_{P}}_{j=1}}{\sum_{j \in N_s}} \sum^{B}_{b=1} \phi_{j, b, t} = 0  \quad \forall \quad t:  \quad \lambda_t \hfill \label{market_balance_cons} \\
    & \added{0 \leq \phi_{j, b, t} \leq \overline{\Phi}_{j, b, t} \qquad \qquad \forall j \in N_s, b \in B, t \hfill \label{bid_block_availability_cons}}
\end{flalign}

The objective defined in Equation~\ref{max_sw_obj} maximizes social welfare defined as gross surplus from satellite bid blocks minus the generation cost incurred by central model generators. Equation~\ref{market_balance_cons} represents the market balance constraint, \replaced{which consists of a set of bid blocks from each prosumer agent and a single bid block per generator agent}{enforcing that the dispatch of central-model generators ($\sum_{i \in N_c} g_{i,t}$) plus the cleared bid blocks from satellite agents ($\sum_{j \in N_s, b} \phi_{j,b,t}$) sum to zero at every time step}. \added{Equation~\ref{bid_block_availability_cons} caps the cleared quantity of each bid block to the maximum quantity $\overline{\Phi}_{j,b,t}$ submitted by satellite agent $j$ for block $b$ at time $t$ to prevent the market from clearing unbounded quantities. By convention, a positive value of $\phi_{j,b,t}$ denotes a supply bid (net injection by satellite agent $j$ into the electric bus) and a negative value denotes a demand bid (net withdrawal); both directions enter the market balance (Equation~\ref{market_balance_cons}) with a consistent sign so that supply bids are matched against demand bids and the residual is balanced by central model generators.}

\subsubsection*{Satellite models}
\replaced{Given an electricity price forecast in scenario $s$, each satellite model solves the following optimization problem.}{Each satellite agent runs a local optimization problem that maximizes its own objective subject to its local technical constraints. The form of this local problem differs between generator-satellites and prosumer-satellites.}

\added{For each generator-satellite $i \in N_{s}^{g}$, given an electricity price forecast $\lambda^{forecast}_{s,t}$ in scenario $s$, the local optimization problem is:}

\added{
\begin{flalign}
    & \left.
    \begin{aligned}
        & {\text{max}} \sum_{t} \Bigl(\lambda^{forecast}_{s,t} - (\alpha_{i} \cdot g_{i,t} + \beta_{i})\Bigr) \cdot g_{i,t} & \\
        & \text{subject to:}  \quad  \text{constraints defined by} \quad \text{Equation}~\ref{gen_dispatch_limit_cons}
    \end{aligned}
    \right\} \quad \forall s, i \in {N}_{s}^{g} \label{generator_satellite_obj}
\end{flalign}
}

\added{Since the focus of our study is on demand-side flexibility, we assume a simple bidding strategy for a generator-satellite: it submits a single bid block per time step at its marginal cost, with bid quantity equal to the optimal dispatch $g^{*}_{i,t}$ obtained from Equation~\ref{generator_satellite_obj}. This is functionally identical to the bidding behavior assumed for central model generators (where the bid takes the form of the quadratic supply bid term in Equation~\ref{max_sw_obj}); the difference is purely architectural---the dispatch is computed inside the satellite's local optimization rather than inside the central market-clearing problem---and serves to illustrate that the framework supports either modeling choice for any generator.} \added{For each prosumer-satellite $j \in N_{s}^{p}$, given an electricity price forecast $\lambda^{forecast}_{s,t}$ in scenario $s$, the local optimization problem is defined mathematically as follows:}

\begin{flalign}
    & \left.
    \begin{aligned}
        & {\text{min}} \replaced{\sum_{c, a, t} - \lambda^{forecast}_{s,t} \cdot p_{j,t} + (\mu_{c, t} \cdot x_{c, a, j, t})}{\sum_{t} \Bigl(- \lambda^{forecast}_{s,t} \cdot p_{j,t}\Bigr) + \sum_{c, a, t} \Bigl(\mu_{c, t} \cdot x_{c, a, j, t}\Bigr)} \\
        & \text{subject to:}  \quad  \text{constraints} \quad \ref{prosumer_electric_power} - \ref{storage_energy_dynamics_cons}, \quad \ref{converter_dispatch_limit_cons} - \ref{flexible_demand_feasible_region}
    \end{aligned}
    \right\} \quad \replaced{\forall s,j \in {N}_{p}}{\forall j \in {N}_{s}^{p}, \text{ for each forecast scenario } s} \added{\label{prosumer_satellite_obj}}
\end{flalign}

Satellite models implicitly account for uncertainty in market prices by generating bids over a range of different price forecast scenarios $s$ rather than a single price-quantity bid pair. This provides some degree of robustness against deviations in forecasts, since bids are not tied to a single deterministic trajectory, but span a range of possible price realizations.

\added{In the rolling-horizon implementation, each satellite agent's local optimization problem is solved repeatedly for consecutive market windows $w$. Within each window $w$, the time index $t$ spans the look-ahead horizon $\mathcal{T}^{w} = \{t_w, t_w+1, \ldots, t_w + L - 1\}$, where $t_w$ is the start of the window and $L$ is the length of the look-ahead window. Satellites with storage and/or flexible demand manage their intertemporal constraints in a look-ahead rolling-horizon manner through handing off states across windows by setting $e^{r,*}_{j, t_w - 1} = e^{r,*}_{j, t^{\mathrm{commit}}_{w-1}}$ for $r \in \{p, q, h\}$ and $* \in \{S, F\}$, where $t^{\mathrm{commit}}_{w-1}$ denotes the final committed time step of the previous market-clearing window. At each rolling-horizon step, only the dispatch corresponding to the committed portion of the current window is fixed, while the remaining future dispatch decisions are recomputed in the subsequent window.} Satellite models can run with a look-ahead window of arbitrary length to optimize for multi-day scheduling. For a given price forecast, the result of the above prosumer-satellite optimization is the power dispatch time series $p_{j,t}$ and not the demand or supply bid functions required by the central model. To generate bid functions consisting of price-quantity pairs as required by the information exchange protocol in this approach, we developed a heuristic optimizer bidding algorithm for prosumer-satellites. The pseudo-code for the optimizer bidding algorithm developed for prosumer-satellites is shown in Algorithm~\ref{bid_generation_algorithm}.

% For generator-satellites (Equation~\ref{generator_satellite_obj}), no intertemporal state needs to be handed off between windows, since the optimization depends only on the forecast price within the current window. We acknowledge that a 24-hour look-ahead is relatively short for prosumer-satellite assets with multi-day flexibility (notably hydrogen storage), and may give rise to end-of-horizon artifacts in which storage is undervalued at the boundary of the look-ahead window because the satellite does not see the longer-term value of stored energy. Such artifacts contribute to the optimality gap of the market-auction approach reported in Figure~\ref{operational_costs_and_electricity_prices} and are part of, but distinct from, the autonomy effects investigated in this study.

\newcommand{\forecast}{\lambda^{forecast}}
\begin{algorithm}
    \caption{Bid generation algorithm for a prosumer agent}
    \label{bid_generation_algorithm}
    \algrenewcommand\algorithmicindent{1.0em}
    \begin{algorithmic}[1]
        \Require Prosumer $j$, time step $t$, CeilingPrice
        \State $\Pi_j \gets \{ \text{CeilingPrice} \}$  \Comment{Initialize Bid Prices}
        \State $\Pi_j \cup \{ \min (\forecast_{t}, \ldots, \forecast_{t+f} ) \forall f \in$ Flexibilities $\}$
        \State $\Pi_j \cup \{ \mu_{c,t} \cdot \replaced{n^r_{a,j}}{\eta^r_{a,j}} \forall c \in C\}$ \Comment{Opportunity costs}
        \ForAll {$\pi_{j,b,t} \in \Pi_j$}
            \State $\forecast_t \gets \pi_{j,b,t}$
            \State $p^\prime_{j,t} \gets$ simulateProsumerWithForecast($\forecast$)
            \State $\phi_{j,b,t} \gets p^\prime_{j,t}$
            \Comment{bid = $(\pi_{j,b,t}, \phi_{j,b,t})$}
        \EndFor
        \State submitBids($B_j$)
    \end{algorithmic}
\end{algorithm}
% \vspace{-3mm}  %%%%%%%%%%%%%%%%%%%%%%%%%%%%%%%%%%%%%%%%%%%%%%%%%%%%%%%%%%%%%%%%%%%%

In lines 1--3, all relevant bid prices are gathered: ceiling price, minimum forecast price per flexibility look-ahead range, and all opportunity costs for using different energy carriers. \added{The ``Flexibilities'' on line~2 of Algorithm~\ref{bid_generation_algorithm} is the set of shifting horizons attached to the prosumer's flexible demand categories, augmented with the full rolling-horizon look-ahead $L$ as an additional entry --- i.e., Flexibilities $=\,\{N : \text{the prosumer has a flexible-demand category with shifting horizon } N\} \cup \{L\}$. For each $f \in$~Flexibilities, the minimum forecast price over the interval $[t,\,t+f]$ is added as a candidate bid price. For example, a prosumer with two flexible demand categories of shifting horizons 4~h and 12~h would have Flexibilities $=\,\{4, 12, 24\}$ (with the third entry being the 24~h rolling-horizon look-ahead), and line~2 of Algorithm~2 would contribute three candidate bid prices: the minimum forecast price over $[t,\,t+4]$, over $[t,\,t+12]$, and over $[t,\,t+24]$.} Once gathered, lines 4--8 determine the prosumer's demand. The forecast price for time $t$ is set to the considered bid price (line 5), the prosumer reports their net electric power demand given that forecast (line 6), which is then set as the quantity of the associated bid (line 7). When that is done for every bid price, the collection of bids can be submitted as a bid function (line 9). \added{To ensure that the submitted bid function is monotonic in price (as required by the central market-clearing model), we sort the resulting $(\pi,\phi)$ pairs by price and merge any non-monotonic segments into a single block (at the higher of the two prices) before submission.}

\added{The bid functions generated by Algorithm~\ref{bid_generation_algorithm} are feasible by construction through four complementary mechanisms: \textit{(i) ceiling-price anchor:} the bid-price set $\Pi_j$ always contains the ceiling price (line~1), so the cumulative dispatch evaluated at the ceiling corresponds to demand that is guaranteed to clear; \textit{(ii) deadline-forced demand:} flexible demand is internally modeled in the satellite optimization as a ``store'' with a lower-bound trajectory, computed from its shifting horizon and enforced through Equations~\ref{flex_demand_store_energy_dynamics_cons} and~\ref{flexible_demand_feasible_region}. As the delivery deadline approaches, any residual flexible demand is therefore forced into the dispatch and bid at ceiling price; \textit{(iii) technically-feasible dispatch:} every per-price dispatch query on line~6 is obtained from the satellite's LP/MILP with all technical constraints active, so each submitted $(\pi,\phi)$ bid pair corresponds to a feasible local dispatch; and \textit{(iv) market-side cap:} Equation~\ref{bid_block_availability_cons} limits the cleared quantity of each bid block to the quantity offered by the prosumer, preventing the market from clearing more power than was submitted. After market clearing, the satellite reconciles its internal state through a deterministic re-solve with the cleared electric power and electricity price fixed.}

The bidding strategy in Algorithm~\ref{bid_generation_algorithm} assumes price-taking (non-strategic) behavior at the satellite level: each prosumer responds to the forecast price as an exogenous input and does not internalize its own market impact. Consequently, strategic bid manipulation and market-power effects are not considered. If market power were the question of interest, Algorithm~\ref{bid_generation_algorithm} could be replaced by a strategic bidding formulation (e.g., a game-theoretic bi-level optimization) and extended to account for uncertainty through stochastic optimization. However, developing advanced ``perfect'' bidding strategies is beyond the scope of this work. Our rationale is that bidding strategies are inherently private to the agents, and thus it is the responsibility of each agent to construct their bids so that their privacy is \replaced{preserved}{enhanced}. The present implementation in Algorithm~\ref{bid_generation_algorithm}, therefore, serves only as a proof-of-concept. Nevertheless, the modularity provided by the model coupling framework allows different bidding strategies to be implemented and evaluated with minimal changes to the coordination architecture and mechanism.

% \added{We also note that the implementation submits the bid function in incremental (step-bid) rather than cumulative form: the dispatch responses obtained at consecutive bid prices in Algorithm~\ref{bid_generation_algorithm} are differenced to produce marginal quantities, and bid blocks with marginal quantity below a small numerical threshold are filtered out before submission.}

\subsection*{The market auction model coupling software}
The market auction model coupling approach is implemented as a new open-source Python-based software Package called Annular~\cite{Dinga_2025_coupling}. Annular is used to couple the central market model to satellite models. The algorithms for each model are implemented generically using Cronian \cite{DohDinga_2025_cronian} \textemdash a companion package to Annular that builds optimization problems from a collection of \texttt{yaml} and \texttt{csv} files for model configuration and data. Annular provides a clear API for the communication and exchange of messages (bids) defined as multi-index tables. Therefore, the central and satellite models only have to be defined in terms of how they should interact (their conduits). Annular takes care of all overhead, such as communication and parallelization tasks, while letting each satellite model contain its own (private) bidding strategy. \added{Similar to Demoses-admm \cite{doh_dinga_2025_price_response}, inter-model communication in Annular is also fully synchronous: the central model waits for bids from all satellites before performing market clearing, and subsequently broadcasts the cleared dispatch to all satellites prior to advancing the rolling-horizon window. Therefore, the central model does not perform clearing on incomplete or stale information, and satellites only update their internal state after market clearing has been completed.} Annular provides synchronized execution of, and communication between models, whether run on a laptop or a cluster system. Additionally, models in Annular can be pointed to different implementations or modeling tools and/or different bidding strategies, making Annular highly flexible and easily extensible.

\section*{Experiment design}\label{sec:case_study_experiment}

\subsection*{Experiment setup}
To evaluate the viability of our proposed model coupling approaches, we benchmark them against co-optimization by comparing agents' dispatch schedules and system costs. To investigate computational efficiency and scalability, we compare the absolute runtime and scaling performance, as well as the peak memory usage, for each approach. This benchmark provides insights into the trade-off between accuracy (preserving agents' autonomy), theoretical optimality, and scalability. We conduct multiple experiments to investigate computational performance by varying the following parameters.

\subsubsection*{Varying problem size}
Problem size can be varied along two dimensions: time (simulation resolution or horizon) and space (number of agents). As a first experiment, we vary problem size while maintaining the same computing infrastructure. We increase problem size by increasing the simulation horizon from 1 day to a week (with hourly resolution) and the number of agents (coupled models) from 30 to about 350. In the temporal dimension, we focus on increasing the simulation horizon. While temporal resolution and horizon represent different modeling choices, both directly affect the total number of time steps and, consequently, the size of the resulting problem. Increasing the horizon, therefore, provides a representative way to capture the growth in computational complexity that would also arise from finer temporal resolutions, as both lead to larger decision spaces and stronger intertemporal coupling. It is also worth noting that while experiments with larger problem sizes could be performed, the selected problem sizes are sufficient to capture scaling trends along temporal and spatial dimensions without exhausting the available computational resources.

% Yet, even in its current state, our work provides a baseline for future studies to evaluate the scalability of different energy system models.

\added{
\subsubsection*{Varying problem complexity}
As a second experiment, we repeat all problem instances from the first experiment with an added complexity by introducing non-convexity through binary decision variables. Specifically, we explicitly prevent the simultaneous charge and discharge of storage assets that might occur when prices are negative. In addition to Equations~\ref{storage_energy_dynamics_cons}, \ref{storage_SOC}, and \ref{storage_dispatch_limit} that model storage, the following constraints are added to the prosumer subproblem:

\begin{flalign}
    & 0 \leq r^{ch}_{j,t} \leq \omega_{j,t} \cdot \overline{S^{r}_{j}}, \qquad \qquad \quad \forall j, t && \label{storage_charge_binary} \\
    & 0 \leq r^{dc}_{j,t} \leq (1 - \omega_{j,t}) \cdot \overline{S^{r}_{j}}, \qquad \quad \forall j, t && \label{storage_discharge_binary} \\
    & \omega_{j,t} \in \{0, 1\}, \qquad \qquad \qquad \qquad \forall j, t && \label{storage_binary}
\end{flalign}

The binary variable $\omega_{j,t}$ takes value 1 when the storage at prosumer $j$ is charging at time $t$ and 0 otherwise, so that Equations~\ref{storage_charge_binary} and~\ref{storage_discharge_binary} together enforce that the storage cannot simultaneously charge and discharge. We adopt storage complementarity as a representative non-convexity (other forms of non-convexity, such as unit commitment binaries and min-up/min-down constraints, can also be introduced) because it provides a minimal yet sufficient test of how each coordination approach behaves once the problem departs from the LP regime.
}

\subsubsection*{Varying computing infrastructure}
The different modeling approaches make use of different coordination architectures, meaning they also differ in how much they can benefit from parallelism in the problem definition. Knowing how much a modeling approach can benefit from parallelism will help inform which trade-off the user or modeler can afford to make. If, for instance, a better time resolution can be achieved by using additional available computing power, that shifts the trade-off from time vs space to time vs computation cost. Naturally, there will also be a limit to how much benefit can be obtained from parallelism, beyond which it will not make sense to deploy the modeling approach on a system with more parallel CPU cores. To give an initial indication of this trade-off, we vary the computational resources used to run the benchmark between the authors' modern laptop with 16 CPU cores, and a more capable 48-core and 128-core server machine\textemdash Snellius \cite{SURF_Snellius}, as a third experiment. This will indicate some actual possible speed-up from parallelism in the different modeling approaches.

\subsection*{Case study and data source}
To use real energy system data for our experiments, we ran PyPSA-Eur model \cite{NEUMANN_2023_Joule} with Germany only for the weather year 2013. \added{The weather year 2013 is widely used as a representative reference year in European energy system studies, and is therefore adopted here for consistency with existing energy system modeling literature \cite{Gotske_2024_NComms, NEUMANN_2023_Joule, Glaum_2026_Joule, Xiong_2025_ERL, Neumann_2025_NComms}.} The collected data include asset capacities (generators, heat pumps, etc.), generator marginal costs, and hourly availability profiles for renewables; hourly demand profiles for electricity, heat, hydrogen, and transport; hourly price profiles for energy carriers such as methane, hydrogen, and biomass. We slightly increase the capacity of storage to reflect a future system with a substantial amount of flexibility.

A typical single node in a PyPSA-Eur model looks like the MIES in Fig~\ref{MIES_fig}, and several of such nodes are interconnected through electricity transmission lines to form a network. Consistent with the copper-plate assumption introduced in the \textit{Methodology} section\textemdash adopted to isolate the coordination layer and reflect zonal electricity market pricing\textemdash we wrote Python scripts to flatten the original network, resulting in a copper-plate model in which all prosumers and generators are connected to a single electric bus. A prosumer is assumed to own all assets that feed into the bus to which it is attached. If there is more than one demand on a bus, the first is assumed to be the base demand (inflexible), while the others are assumed to be flexible. To vary the number of agents for our scaling experiments, we use PyPSA-Eur clustering functionality to increase the spatial granularity of the network to generate the required number of agents. The smallest network instance (with a spatial granularity of 1) results in about 30 agents, of which 16 are generators and 14 are prosumers. Increasing the spatial granularity by a step size of 1 increases the number of agents by roughly 10, and the proportion of distribution between agent types is constant; for every 10 agents added, about 6 are generators and 4 are prosumers.

\added{The partition between the central model and satellite agents in the market auction experiments is defined as follows: all prosumers are modeled as satellites ($N_{s}^{p} = N_{p}$) for all problem instances; $|N_{g}| - 1$ generators are included in the central model while the remaining generator is modeled as a satellite ($N_{s}^{g} = 1$), submitting a single bid corresponding to its available capacity at marginal cost via the generator-satellite formulation in Equation~\ref{generator_satellite_obj}. The reason for such a partition is twofold: first, it ensures that the focus of the study remains on demand-side flexibility, which is the primary object of analysis; second, it demonstrates the flexibility of the proposed framework, as generators can either be represented as satellite agents or be absorbed into the central model, with both configurations seamlessly supported within Annular.}

We used Gurobi solver and set \texttt{Threads=-1} and \texttt{Method=2} so that Gurobi automatically uses the maximum number of threads for maximum parallelization. For the price-response method, we set a common convergence tolerance $\epsilon = 0.1$, applied to both the primal $\Psi^{k}$ and the dual $\widetilde{\Psi}^{k}$ stopping criteria, and employ the dynamic penalty parameter update strategy of Step~6 of Algorithm~\ref{price_response_admm} to keep the two residuals of comparable magnitude during the iterative process, thereby enabling the use of a single convergence threshold and improving convergence speed. For the market auction method, satellite models are assigned a look-ahead horizon of 24 hours, while the market clearing horizon is set to 1 hour. A 24-hour look-ahead is relatively short for assets with multi-day flexibility, such as hydrogen storage, and may give rise to end-of-horizon artifacts. In such cases, storage can be undervalued near the boundary of the look-ahead window because the satellite optimizer does not fully capture the longer-term value of stored energy beyond the optimization horizon. This limited foresight is expected to contribute to the optimality gap of the market auction approach. Nevertheless, to quantify the magnitude of this effect, we perform a sensitivity analysis in which the look-ahead window is extended, and its impact on the optimality gap is evaluated.

\section*{Results and Discussion}

\subsection*{Accuracy vs. Optimality}
This section compares the performance of the three coordination approaches in terms of optimality and shows the impact of preserving agents' autonomy (more realistic representation of agent behavior) on the "ideal" optimum computed by co-optimization. We first present the dispatch schedules of agents under each method, and subsequently show the total system operational costs. Finally, we conclude the section by presenting results from a detailed decomposition and sensitivity analysis of the optimality gap between co-optimization and the two model coupling approaches.

\subsubsection*{Dispatch schedules of agents}
\begin{figure*}[htbp]
    \centering
    \captionsetup{justification=centering}

    \begin{subfigure}[b]{\textwidth}
        \centering
        \caption{Aggregate dispatch of all electricity generators}
        \includegraphics[width=\textwidth]{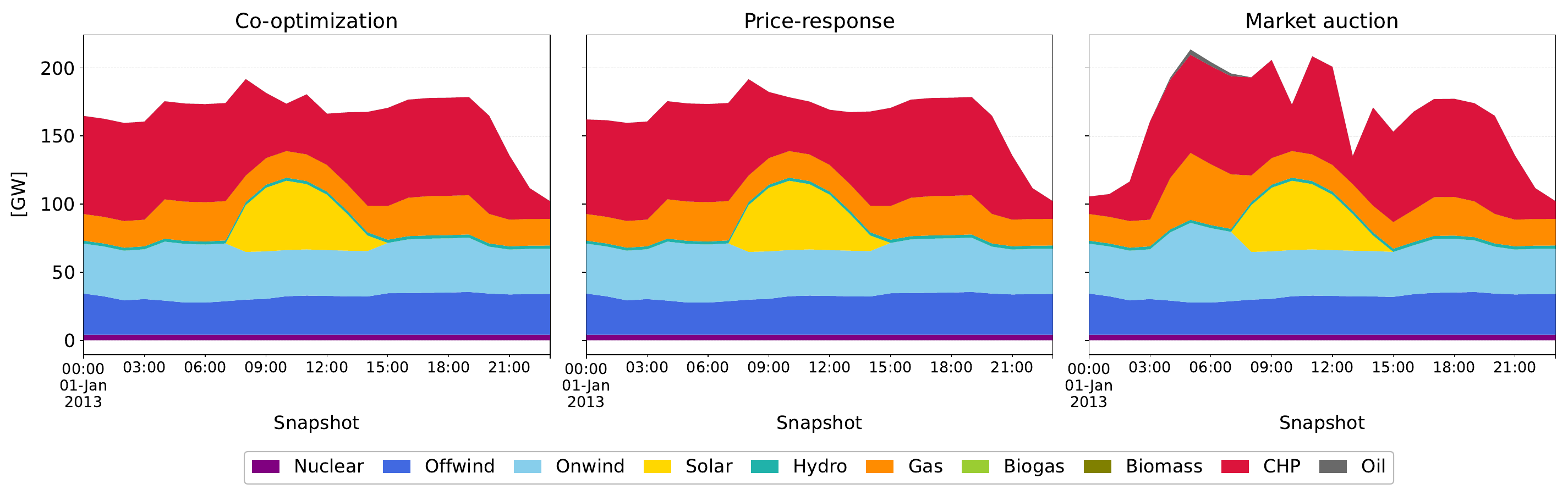}
        \label{generator_dispatch}
    \end{subfigure}

    \begin{subfigure}[b]{\textwidth}
        \centering
        \caption{Aggregate dispatch of all battery storage prosumers}
         \hspace*{0.8em}
        \includegraphics[width=\textwidth]{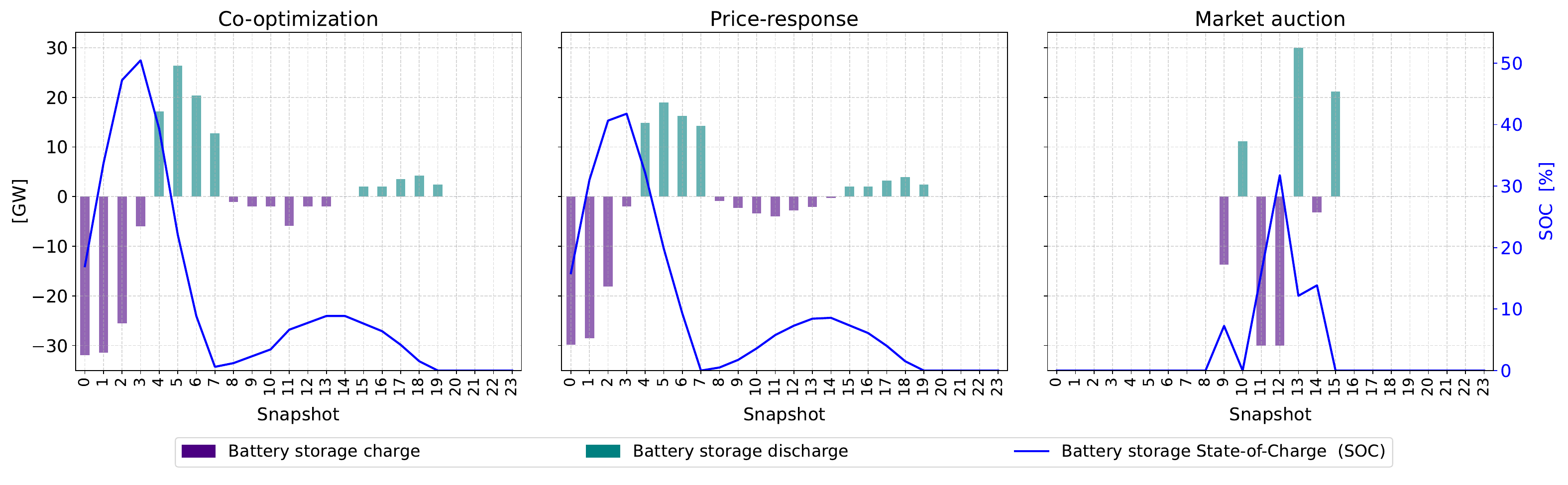}
        \label{battery_dispatch}
    \end{subfigure}

    \begin{subfigure}[b]{\textwidth}
        \centering
        \caption{Aggregated dispatch across all heat subsystem prosumers}
        \includegraphics[width=\textwidth]{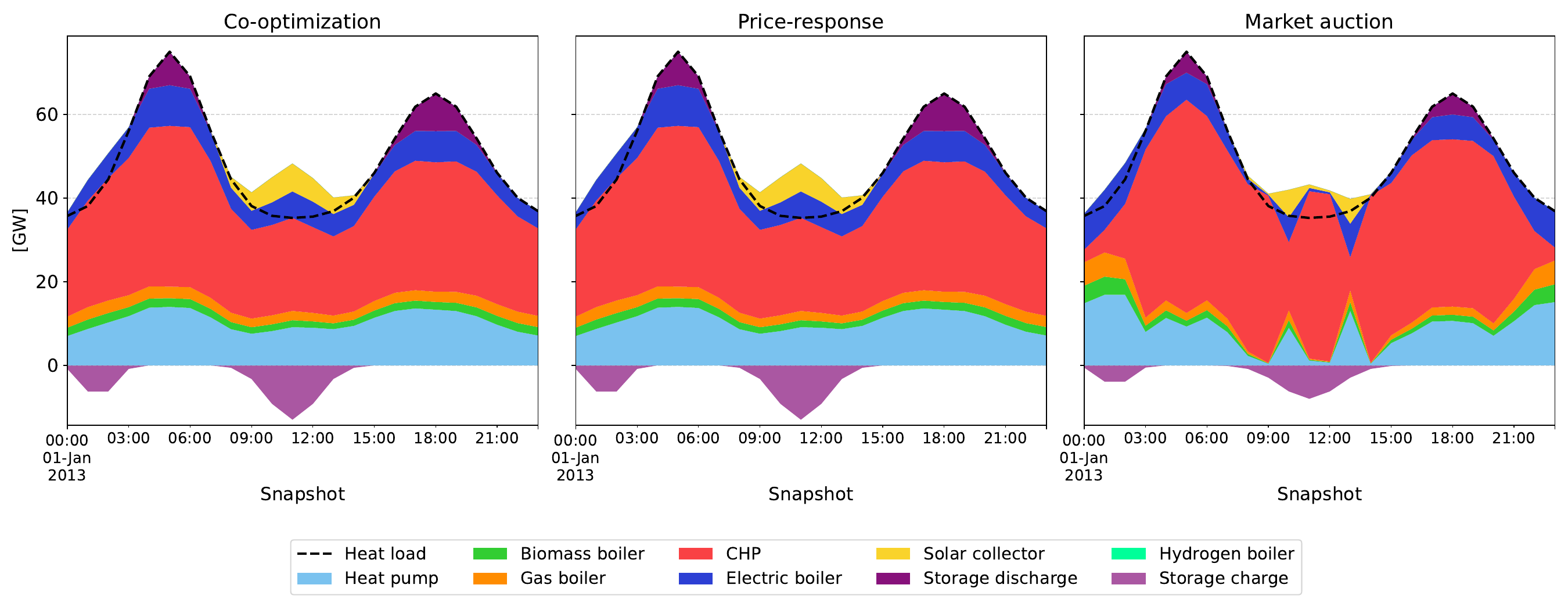}
        \label{heat_prosumer_dispatch}
    \end{subfigure}

    \caption{Dispatch schedules under the three coordination approaches for the smallest problem instance: (a) aggregate generator dispatch; (b) aggregate battery storage dispatch; (c) aggregate heat subsystem dispatch.}
    \label{dispatch_schedules}
\end{figure*}

The dispatch schedules for all generators, battery storage, and heat prosumers in the experiment with the smallest problem instance are shown in Figure~\ref{dispatch_schedules}. As expected, the dispatch schedules of agents from co-optimization and price-response are almost identical (see Figure~\ref{optimality_gap_decomposition} on the decomposition of the optimality gap). However, the difference in agents' dispatch between market auction and the other two approaches is more pronounced. Take, for instance, the dispatch of electricity generators; a significant difference can be observed in how CHP, gas, and oil-fired generators are used to satisfy peak demand. In the market auction approach, less power is produced by these three generators in earlier periods. However, there is a steep ramp-up in power from these generators thereafter. This difference can be explained through the battery storage dispatch in Figure~\ref{battery_dispatch}. Due to its perfect foresight, full knowledge, and top-down controllability of the MIES, co-optimization can anticipate high prices in the future, and therefore dispatch more production from CHP and gas at an earlier time to charge the battery. The battery is then used to supply peak demand later, thereby preventing the more expensive oil-fired generator from being dispatched at all. The same trend can be observed for the price-response approach, which achieves a very similar dispatch schedule as co-optimization through several iterations.

Conversely, the battery schedule under the market auction approach differs substantially. This difference arises from agents' autonomy and myopic decision-making under imperfect information. Because battery operators retain full autonomy in market auction, they optimize their dispatch based on local objectives, namely revenue maximization. As described in Algorithm~\ref{bid_generation_algorithm}, the battery operator identifies the lowest forecast price within the horizon and places buy bids at that time in order to discharge at higher prices later. Since the lowest forecast price occurs later in the horizon, this leads to delayed charging and therefore a failure to arbitrage earlier price peaks. As a consequence, early peak demand is instead met by gas, CHP, and oil-fired generators. This behavior reflects the price-taking and imperfect forward-looking limitations of the bidding heuristic in Algorithm~\ref{bid_generation_algorithm}, which does not internalize market impact or anticipate system-level effects. Therefore, the resulting dispatch differs from the perfect foresight co-optimization benchmark, which can coordinate intertemporal arbitrage optimally by construction. This difference between co-optimization and market auction reflects the characteristic of real-world electricity markets that arise from decentralized decision making and forecast-driven operations under limited foresight.

% \deleted{Therefore, unlike co-optimization and price-response, the market auction approach captures decentralized market behavior and thereby enables the study of system dynamics as they arise in practice, providing a modeling environment for testing policy and market design.}

% On the other hand, the \replaced{aggregate dispatch across all heat prosumers shown in Figure~\ref{heat_prosumer_dispatch}}{dispatch of a heat prosumer shown in Figure~\ref{heat_prosumer_dispatch}} is very similar for all three approaches. While the dispatch in co-optimization and price-response are almost identical, the only significant difference between these and the market auction method is in the operational schedule of the electric boiler and heat storage. In the market auction method, the electric boiler is not dispatched until around 03:00 (because there is limited generation from gas). This means there is less heat to buffer, hence, a lower amount of heat is stored in the heat storage.

\subsubsection*{Total operational costs and electricity prices}
\begin{figure*}[htbp]
    \centering
    \captionsetup{justification=centering}
    \begin{subfigure}[b]{0.49\textwidth}
        \centering
        \caption{Total operational costs}
        \includegraphics[width=\textwidth]{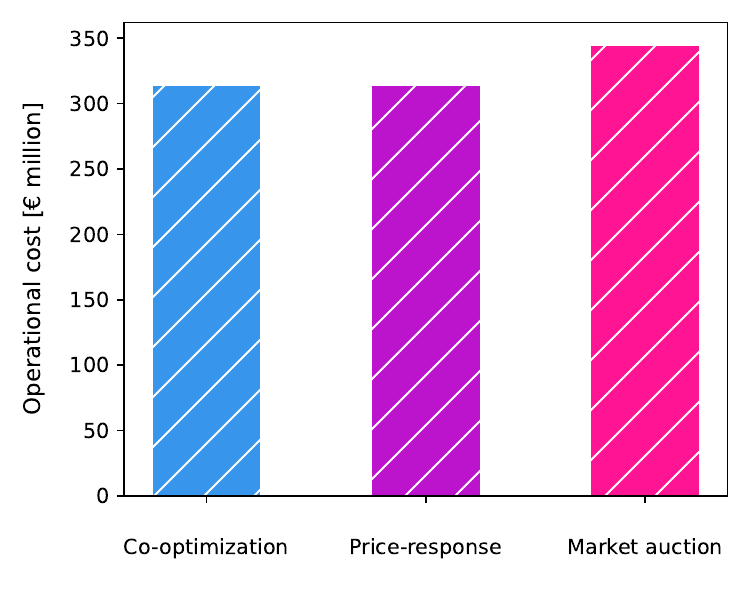}
        \label{operational_cost}
    \end{subfigure}
    \hfill
    \begin{subfigure}[b]{0.49\textwidth}
        \centering
        \caption{Electricity prices}
        \includegraphics[width=\textwidth]{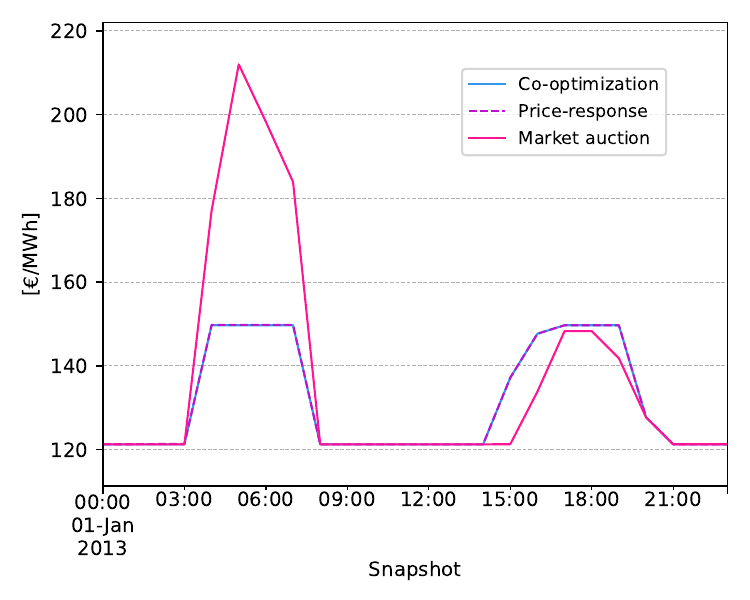}
        \label{market_prices}
    \end{subfigure}
    \caption{Total operational cost and electricity price for the smallest problem instance under the three coordination approaches: (a) total operational cost of the MIES; (b) hourly market-clearing electricity prices.}
    \label{operational_costs_and_electricity_prices}
\end{figure*}

The total operational cost of the MIES and electricity prices are shown in Figure~\ref{operational_costs_and_electricity_prices}. The total operational cost (see Figure~\ref{operational_cost}) under the three approaches is around €314.2, €314.3, and €344.8 million for co-optimization, price-response, and market auction, respectively (see Figure~\ref{optimality_gap_decomposition} for a detailed analysis of this difference). \added{The higher operational cost under the market auction approach arises directly from the decentralized and forecast-driven dispatch behavior discussed above. In particular, the inability of the battery operator to anticipate and arbitrage the early-morning scarcity period forces the system to rely more heavily on expensive thermal generation.} \added{
The electricity-price trajectories are also shown in Figure~\ref{market_prices}. Co-optimization and price-response produce nearly identical prices, whereas the market auction approach exhibits a similar overall trend, but with substantially sharper price spikes, most notably around~04:00. This spike is directly linked to the battery dispatch behavior discussed above. Because the battery operator delays charging and leaves the battery idle during the early-morning peak (see Figure~\ref{battery_dispatch}), the system lacks low-cost stored energy that could otherwise serve the additional demand. The residual demand must therefore be supplied by the oil-fired generator, which becomes the marginal unit during that hour. Since market-clearing prices are set by the marginal generator, the dispatch of the oil-fired unit directly causes the electricity price spike to approximately~210~€/MWh in Figure~\ref{market_prices}.
} 

% The same causal chain of events: battery inactivity $\rightarrow$ expensive generator dispatch $\rightarrow$ price spike explains the other scarcity spikes observed in the market auction electricity price profile.

It is worth noting that the costs and prices obtained from co-optimization should only be interpreted as a theoretical reference benchmark and not as a strict performance metric, since they result from an ``ideal'' setting characterized by perfect foresight, perfect competition, and top-down centralized dispatch; whereas actual market outcomes are almost never optimal in reality. Accordingly, the objective of the market auction approach is not to minimize prices directly (see Figure~\ref{decision_tree} on the nuance in model selection), but rather to provide a modeling environment capable of fully representing the behavior and interactions of flexibility providers. This, in turn, provides a testbed for evaluating how alternative policy regulations and new market design mechanisms influence the behavior of flexibility providers and their resulting impacts on system-wide social welfare.

In sum, our findings show that while both co-optimization and price-response can find the theoretical system-wide optimum, they are, however, less accurate (with respect to modeling flexibility under imperfect competition) since both methods disregard features such as \replaced{strategic behavior}{the myopic, bounded-rationality bidding behavior of self-interested flexibility providers} that are inherent in real-world decentralized market settings. On the other hand, while market auction does not achieve the same "ideal" optimum as the other two approaches, it can fully capture these key features of flexibility that must be studied and understood\textemdash in order to avoid myopic operational and investment decisions for market parties, and for regulators to investigate policy instruments that can lead to an efficient integration and utilization of flexibility resources in MIES.

% \deleted{This highlights the difference between the ``ideal'' system-wide optimum computed by co-optimization and the higher, yet more realistic, costs that arise from the full autonomy of self-interested agents whose \replaced{strategic objectives optimize local revenue}{ price-taking bidding heuristics optimize against an imperfect price forecast} rather than the system-wide optimum. The main driver of the cost difference\textemdash electricity prices\textemdash is shown in Figure~\ref{market_prices}.}

\added{
\subsubsection*{Decomposition and sensitivity analysis of the optimality gap}
The total operational costs of the MIES reported in Figure~\ref{operational_costs_and_electricity_prices} correspond to the smallest problem instance (30 agents, 24~h simulation horizon) under the default settings of the market auction approach: a 24~h look-ahead window and a step-function bid resolution of 5 price--quantity blocks. Under these settings, the market auction approach exhibits an approximately 9.7\% higher operational cost than co-optimization, which can be interpreted as the ``theoretical'' optimality gap between the two approaches. However, two questions remain unanswered: (i)~does this gap persist across different problem instances? and (ii)~which modeling choices drive the gap, and to what extent?

To address these questions, we perform a decomposition and sensitivity analysis of the objective-cost deviation across all 44 problem instances (4 simulation horizons~$\times$~11 agent counts). Figure~\ref{optimality_gap_decomposition} reports the optimality gap (percentage deviation in total operational cost relative to co-optimization) under three experimental conditions: (i)~the baseline configuration (24~h look-ahead, 5 bid blocks); (ii)~a higher bid resolution variant (10 bid blocks, look-ahead unchanged); and (iii)~a longer look-ahead variant (168~h look-ahead, bid resolution unchanged). These two sensitivity dimensions apply only to the market auction approach and therefore do not affect the price-response results; consequently, the left-column panels in Figure~\ref{optimality_gap_decomposition} are identical across all three rows, and only serve to show the optimality gap across different problem instances in the price-response approach.

\begin{figure*}[htbp]
    \centering
    \includegraphics[width=0.95\textwidth]{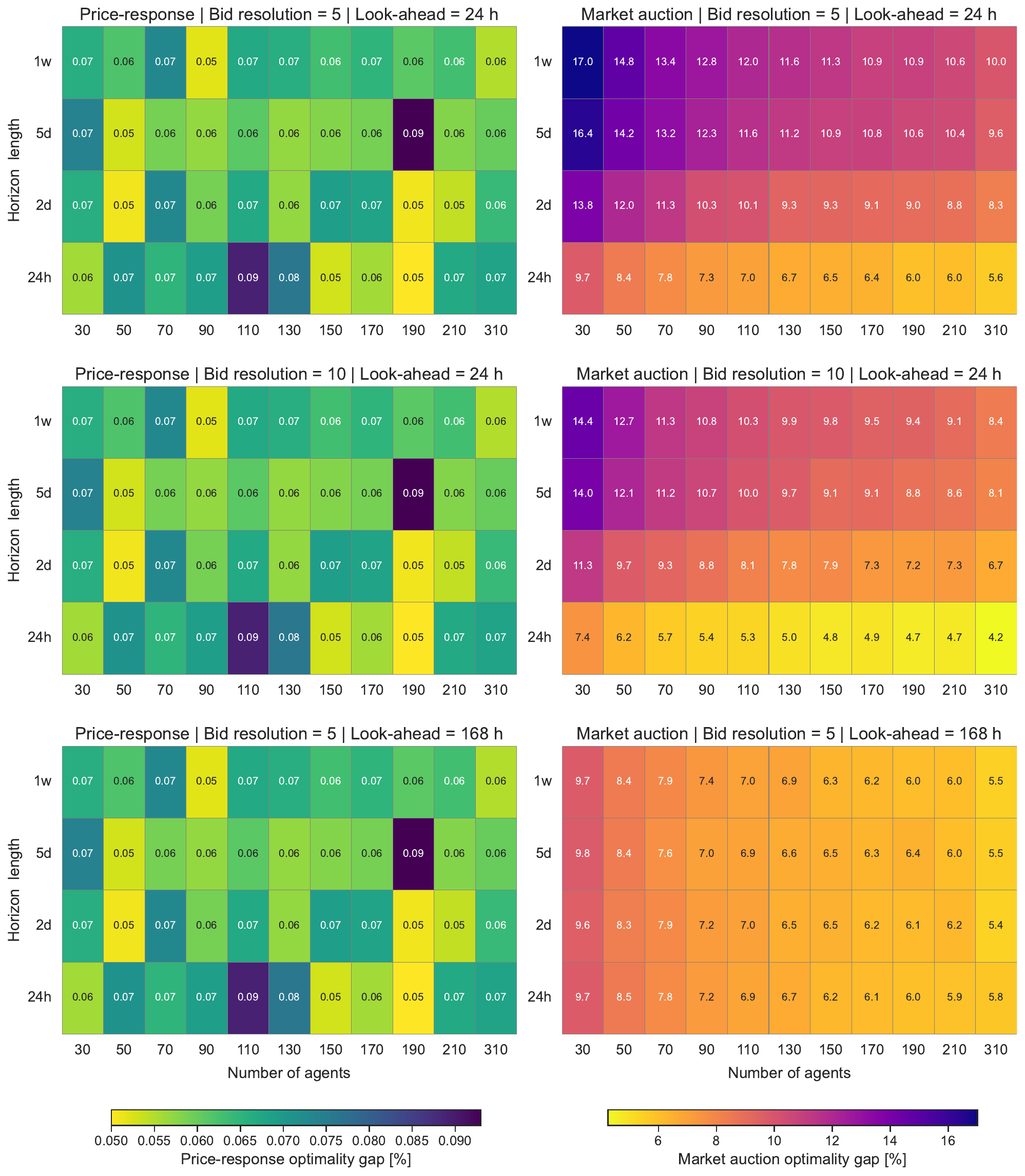}
    \caption{Decomposition and sensitivity analysis of the optimality gap (\% deviation in total operational cost from co-optimization) for the price-response (left column) and market auction (right column) approaches, across all 44 convex problem instances (4 simulation horizons $\times$ 11 agent counts) under three experimental conditions (rows): a baseline (look-ahead = 24~h, bid resolution = 5 blocks); a higher-resolution bid variant (10 blocks, look-ahead unchanged); and a higher-look-ahead variant (168~h, bid resolution unchanged). Cells are annotated with the gap value. The two variants are knobs of the market auction approach and do not apply to the price-response approach: the price-response panels (left column) are therefore identical across all three rows and simply report the price-response optimality gap across instances; they are repeated in rows~2 and~3 only to allow a per-row visual comparison against the market auction panel on the right.}
    \label{optimality_gap_decomposition}
\end{figure*}

Two observations stand out. First, the price-response approach (left column of Figure~\ref{optimality_gap_decomposition}) exhibits an optimality gap below 0.1\% across all 44 problem instances. The results are identical across the three rows because the price-response approach does not rely on rolling-horizon look-ahead or bid functions, and is therefore unaffected by either sensitivity dimension. This empirically confirms the theoretical expectation that price-response recovers the centralized co-optimization solution within the predefined convergence tolerance for convex problems. Second, the market auction approach (right column of Figure~\ref{optimality_gap_decomposition}) exhibits a substantially larger and more variable gap, which can be decomposed into the following three main effects.

(i) \textit{Foresight effect.} Extending the look-ahead window from 24~h to 168~h reduces the gap in the worst-case instance (30 agents, 1-week simulation horizon) from approximately 17\% to 9.7\%, corresponding to a reduction of about 7 percentage points. This quantifies the effect of imperfect foresight. With a 24~h look-ahead applied to a 168~h simulation, satellites optimize only over a limited future window at each commitment step. When the look-ahead is extended to cover the full simulation horizon, satellites effectively operate under perfect foresight and this component of the gap largely disappears.

(ii) \textit{Bid resolution effect.} Increasing the bid resolution from 5 to 10 price--quantity blocks reduces the gap in the baseline instance (30 agents, 24~h horizon) from approximately 9.7\% to 7.4\%, a reduction of about 2.3 percentage points that is relatively consistent across instances. This reflects the approximation error introduced by representing the satellite's underlying cost curve with a coarse step-function bid. A higher bid resolution provides a more granular representation of agents’ underlying marginal cost and willingness-to-pay curve. This allows the market to reflect finer variations in marginal valuation that would otherwise be aggregated within coarser bid blocks, leading to a more efficient market-clearing outcome.

(iii) \textit{Agent averaging effect.} Across all market auction panels, the optimality gap decreases as the number of agents increases. Under the baseline configuration, the gap falls from approximately 9.7\% at 30 agents to 5.6\% at 310 agents, while under the higher bid resolution configuration it decreases from 7.4\% to 4.2\%. This reflects an averaging effect: as the number of participating agents grows, the individual imperfections and myopia of local bidding heuristics become partially averaged out through aggregate market clearing, resulting in more efficient system-level outcomes.

Together, these three effects explain most of the variation in the market auction optimality gap. The worst-case configuration (30 agents, 1-week horizon, 24~h look-ahead, 5 bid blocks) reaches approximately 17\%, whereas the best-case configuration (310 agents, 24~h horizon, 24~h look-ahead, 10 bid blocks) falls to approximately 4.2\%. The remaining 4--5\% gap, therefore, represents the residual efficiency cost inherent to decentralization, even under favorable modeling assumptions. Notably, this residual range is comparable to the 1--5\% efficiency-loss range commonly reported for well-functioning wholesale electricity markets~\cite{Borenstein_2002_AER, Cramton_2017_OREP}, suggesting that the efficiency cost of fully decentralized coordination is of similar magnitude to inefficiencies already observed in decentralized electricity markets.
}

\subsection*{Computational efficiency and scalability}
In this section, we compare the computational efficiency (reflected by absolute runtime) and scalability (reflected by normalized runtime) of the three coordination approaches under convex and non-convex problem instances. We also discuss the runtime composition (the time spent on computation, communication, and waiting) of price-response and market auction. Finally, we conclude with an analysis of peak memory usage (reflected by Maximum Resident Set Size, RSS) across the three approaches.

\subsubsection*{Runtime and scalability}
\begin{figure*}[htbp]
    \centering
    \captionsetup{justification=centering}

    \begin{subfigure}[b]{\textwidth}
        \centering
        \caption{Cumulative absolute runtime}
        \includegraphics[height=0.325\textwidth]{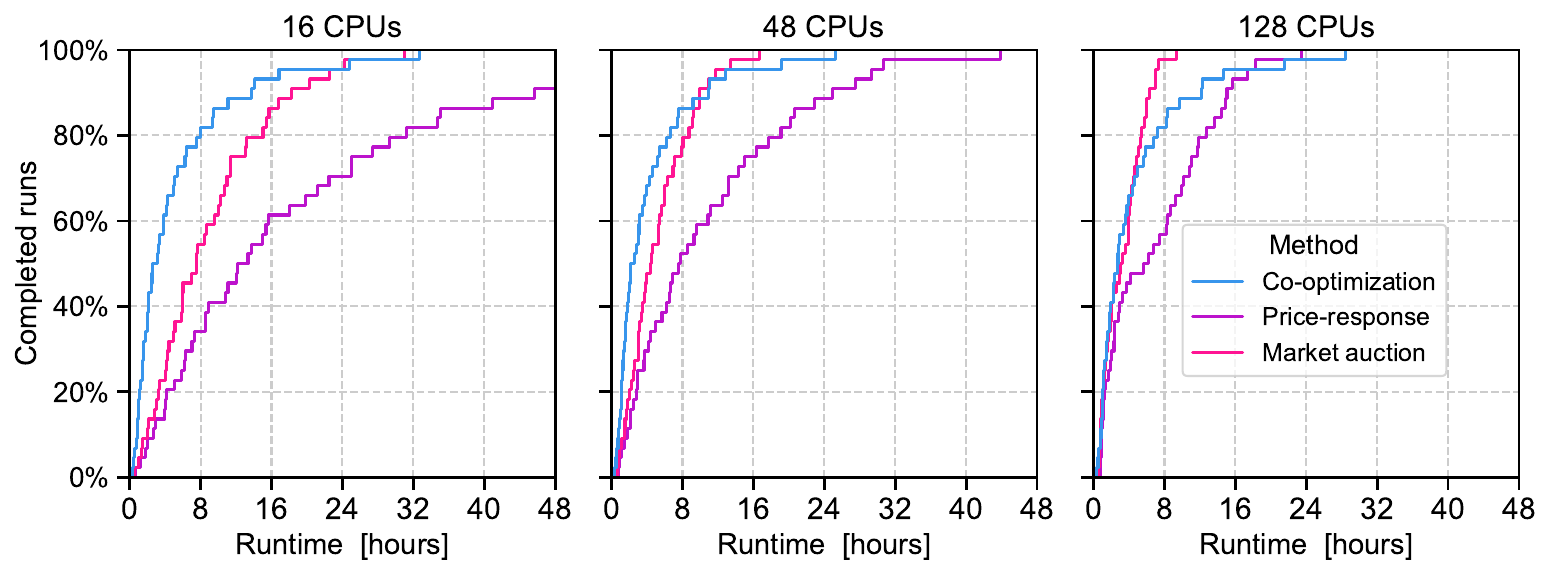}
        \label{ecdf_absolute_runtime_convex}
    \end{subfigure}

    % \vspace{-1em}  % Optional: add vertical space between the subfigures

    \begin{subfigure}[b]{\textwidth}
        \centering
        \caption{Normalized runtime per experiment}
        \includegraphics[height=0.88\textwidth]{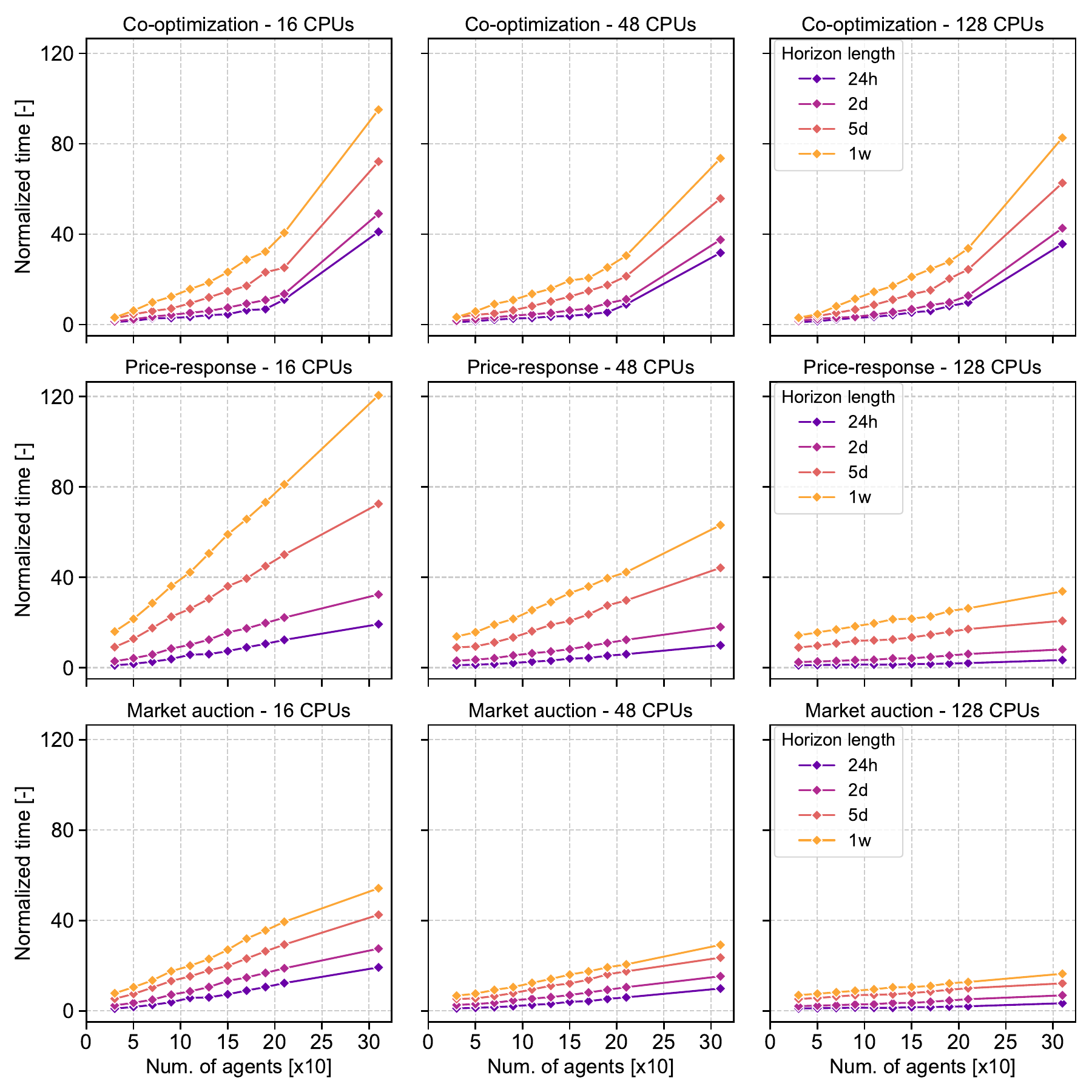}
        \label{scaling_results_convex}
    \end{subfigure}

    \caption{Scaling performance of the three coordination approaches across the 44 convex (LP) problem instances (4 simulation horizons × 11 agent counts), run at three CPU configurations (16, 48, 128 cores): (a) empirical cumulative distribution of wall-clock runtime; (b) per-method minimum-time-normalized runtime.}
    \label{convex_runtime_and_scaling}
\end{figure*}

The computational efficiency and scalability results for the convex problem instances are shown in Figure~\ref{convex_runtime_and_scaling}. We start with the computational efficiency reflected by the cumulative distribution of absolute runtime across all convex problem instances (Figure~\ref{ecdf_absolute_runtime_convex}). Each problem instance\textemdash defined by a unique combination of number of agents and horizon length\textemdash is run as a separate job. Therefore, Figure~\ref{ecdf_absolute_runtime_convex} conveys two types of information: (i) the total time required to complete all problem instances under different experimental settings for each approach; and (ii) the speed of computations, expressed as the proportion of solved problem instances as a function of wall-clock time.

% ====================================== convex_runtime_and_scaling ==========================================
% ================================================================================================================

The following observations can be made from Figure~\ref{ecdf_absolute_runtime_convex}. First, the market auction approach is the fastest overall, as indicated by its lowest cumulative absolute runtime across all experiments. Price-response outperforms co-optimization only when more CPU cores are available. \added{The practical implication is that the relative ranking of the three approaches is hardware-dependent: on a 16-core machine (the leftmost panel of Figure~\ref{ecdf_absolute_runtime_convex}), the price-response approach is in fact the slowest of the three across most of the problem instance distribution, and only becomes competitive with co-optimization once more CPU cores are available to parallelize the per-agent subproblems.} From the steep initial rise of its cumulative distribution curve, co-optimization is fastest for small problem instances, as it avoids coordination overhead associated with iterative updates and communication overhead among models. However, the runtime of co-optimization increases sharply as problem size grows, leading to significantly slower performance for larger problem instances. Therefore, while co-optimization is efficient for small-scale problems, it becomes computationally inefficient as the problem size increases.

Second, increasing the number of CPU cores results in significant speedup (lower runtime) for both the price-response and market auction approaches, as reflected by the leftward shift of their curves across subplots. This effect is most pronounced for the market auction approach; market auction improves cumulative batch throughput by approximately 30\% on the 16-core configuration, and by up to 70\% on the 128-core HPC node relative to co-optimization. \added{These gains are batch-throughput gains across all problem instances rather than per-instance speed-ups; the per-instance speed-up at the 128-core level is approximately a factor of 2-3 for larger problem sizes.} For co-optimization, an initial increase in the number of CPU cores from 16 to 48 leads to a moderate speedup. However, further increasing the number of CPU cores from 48 to 128 actually results in longer co-optimization runtime due to diminishing returns at higher parallel thread counts. That is, additional threads duplicate the working memory, and the per-thread synchronization overhead eventually offsets parallelization gains. This highlights the extent to which different coordination architectures can efficiently exploit parallel computing resources: the distributed and decentralized architectures of model coupling scale naturally with additional cores because they decompose the problem into many smaller subproblems that can be solved concurrently, whereas the monolithic co-optimization formulation exhibits limited parallel scalability due to the centralized structure of the problem definition.

To further assess the scaling trend of each approach, we normalize the runtime of each problem instance by the runtime of the smallest instance for each approach. Normalization also eliminates any bias from implementation-specific optimizations, since each method is compared against its own baseline. The result of the scaling trends for convex problem instances is shown in Figure~\ref{scaling_results_convex}. We see that under fixed computational resources, the solution time for co-optimization increases rapidly with problem size at \replaced{a roughly exponential rate}{an approximately polynomial rate ($O(n^2)$ in the agent count $n$), consistent with the typical behavior of barrier and simplex solver methods on this class of problems}, whereas both price-response and market auction scale almost linearly. Among these, the market auction approach scales substantially faster as reflected by its lowest slope. Consistent with the cumulative distribution of absolute runtime (see Figure~\ref{ecdf_absolute_runtime_convex}) discussed earlier, we again see that as computational resources increase, co-optimization shows a moderate improvement in speedup up to 48-cores, and then slows down as the number of CPU cores increases further. On the other hand, price-response and market auction benefit significantly from parallelism, as indicated by the reduced slopes of their curves across subplots.

\begin{figure*}[htbp]
    \centering
    \captionsetup{justification=centering}

    \begin{subfigure}[b]{0.49\textwidth}
        \centering
        \caption{Convergence of price-response on a representative convex instance}
        \includegraphics[width=\textwidth]{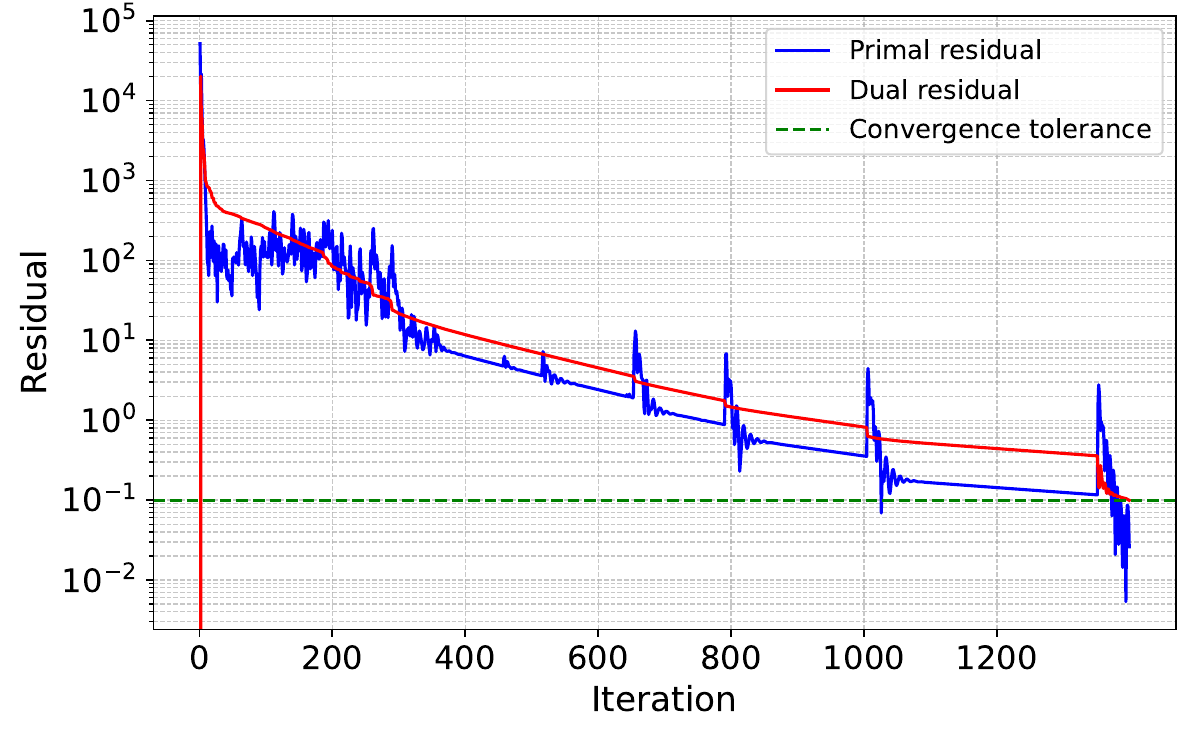}
        \label{primal_dual_residuals_convex}
    \end{subfigure}
    \hfill
    \begin{subfigure}[b]{0.49\textwidth}
        \centering
        \caption{Divergence of price-response on a representative non-convex instance}
        \includegraphics[width=\textwidth]{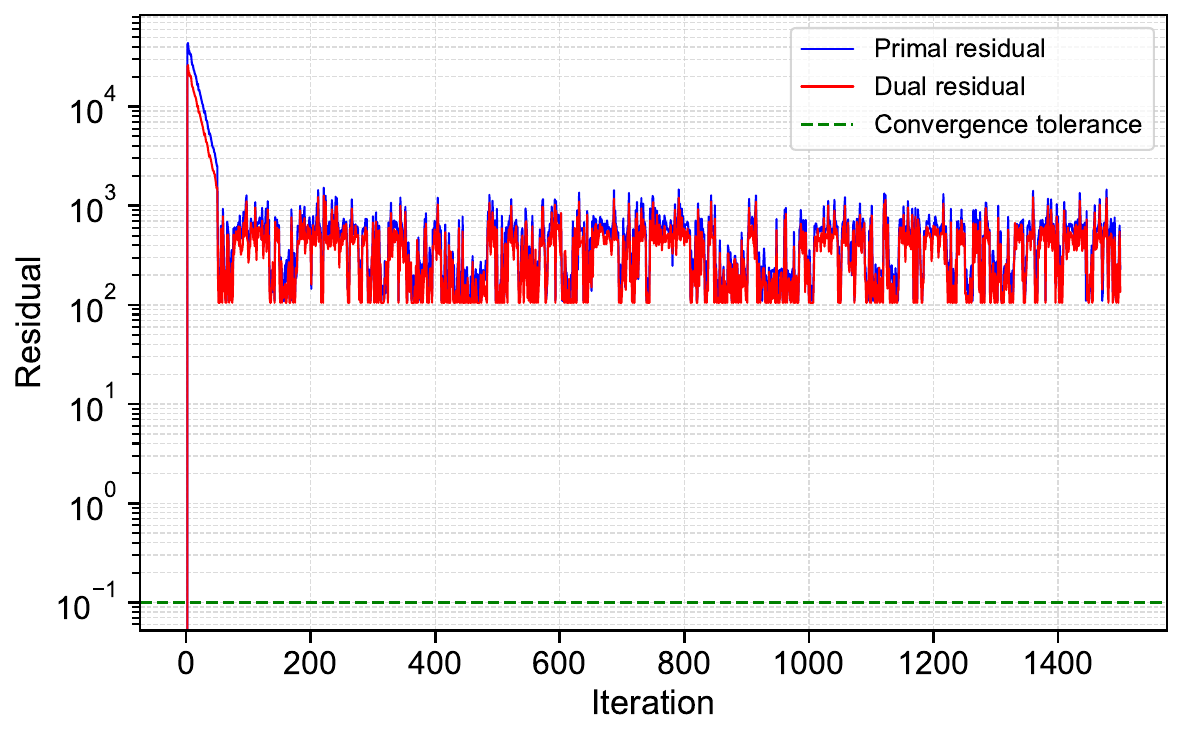}
        \label{primal_dual_residuals_nonconvex}
    \end{subfigure}

    \begin{subfigure}[b]{\textwidth}
        \centering
        \caption{Iteration count to convergence across all convex problem instances}
        \includegraphics[width=0.63\textwidth]{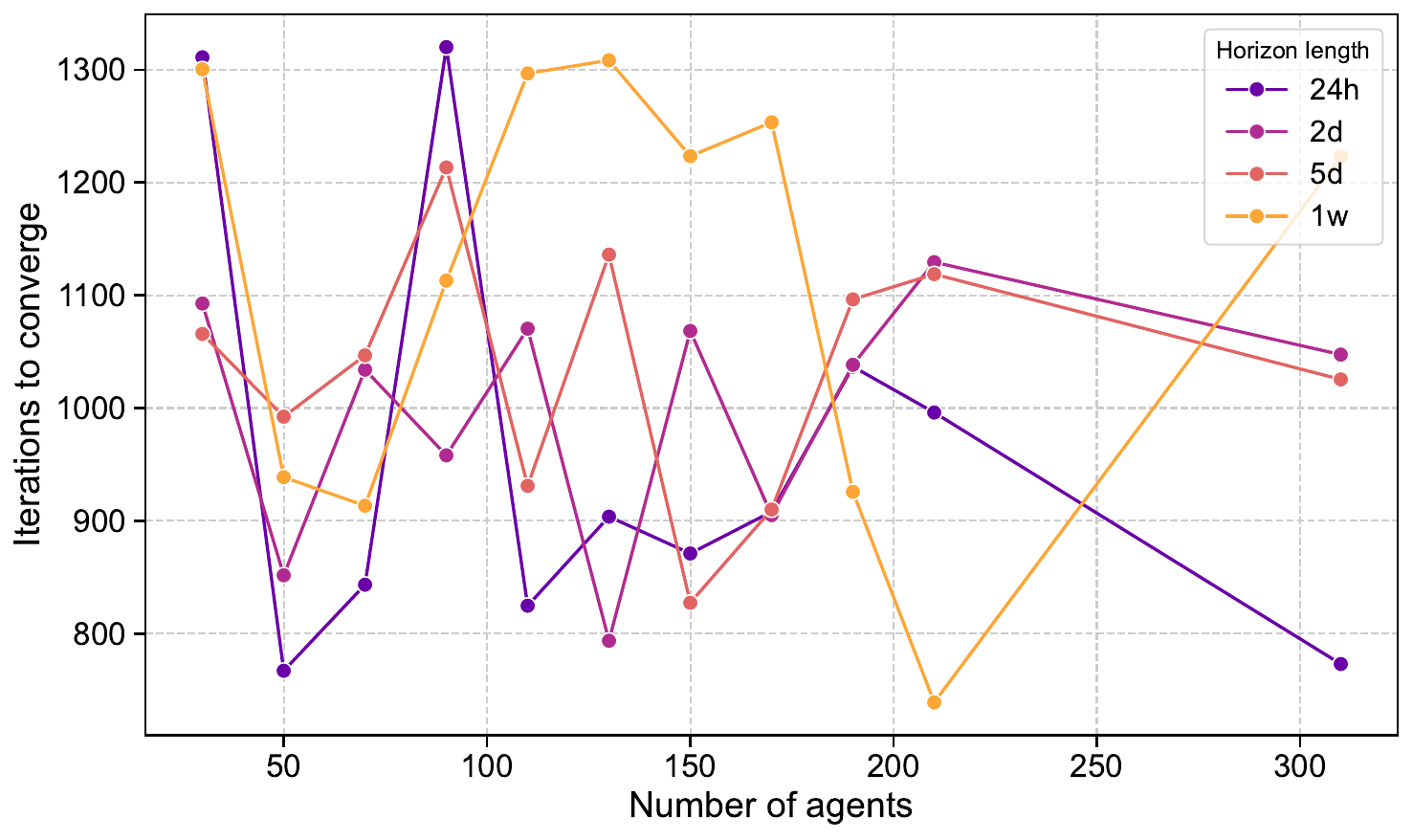}
        \label{iteration_count_vs_size}
    \end{subfigure}

    \caption{Analysis of the convergence behavior of price-response. (a) Primal and dual residuals decay to the convergence tolerance for a representative convex instance. (b) On a representative non-convex instance, both residuals enter a sustained flip-flop state after an initial fast decay and never cross the tolerance. (c) Iteration count to convergence across all 44 convex problem instances (4~horizons~$\times$~11~agent counts); the mean is ~1000, with no monotonic scaling trend in either dimension.}
    \label{pr_convergence_behavior}
\end{figure*}

The difference in scalability can be explained by differences in the coordination architecture, which determines how efficiently each modeling approach can exploit parallelism (as discussed above), and the coordination mechanism, which determines the number of computational steps required per solution. Co-optimization exhibits the weakest scaling performance since it relies on a monolithic problem structure that offers limited opportunities for decomposition and parallel execution. The market auction approach is substantially faster than the price-response approach because it is non-iterative---that is, the interaction between the central and satellite models is executed once per market-clearing round. In contrast, the iterative nature of price-response introduces a computational bottleneck since agents must repeatedly solve their local flexibility scheduling problems until convergence (see Figure~\ref{primal_dual_residuals_convex} and Figure~\ref{primal_dual_residuals_nonconvex}). This implies that the total solution time of price-response  increases proportionally to the number of iterations required for convergence. Moreover, while price-response only decomposes the problem across space (agents), market auction decomposes the problem across both space (agents) and time, since it partitions the horizon into smaller rolling windows and solves them sequentially in a rolling manner.

\added{We additionally examine whether the iteration count of the price-response scales with problem size. Figure~\ref{iteration_count_vs_size} reports the iteration count to convergence across all 44 convex problem instances. The mean iteration count is about 1000, with per-instance counts ranging from approximately 700 to 1300 and showing no monotonic trend with respect to either the agent count or horizon length. Therefore, the convergence behavior of the price response is dominated by its consensus dynamics rather than the dimensionality of the underlying problem, implying that larger instances do not necessarily require more iterations but instead require more computational effort per iteration.}

% ======================== Analysis of the convergence behavior of price-response ================================
% ================================================================================================================

\begin{figure*}[htbp]
    \centering
    \captionsetup{justification=centering}

    \begin{subfigure}[b]{\textwidth}
        \centering
        \caption{Cumulative absolute runtime}
        \includegraphics[height=0.325\textwidth]{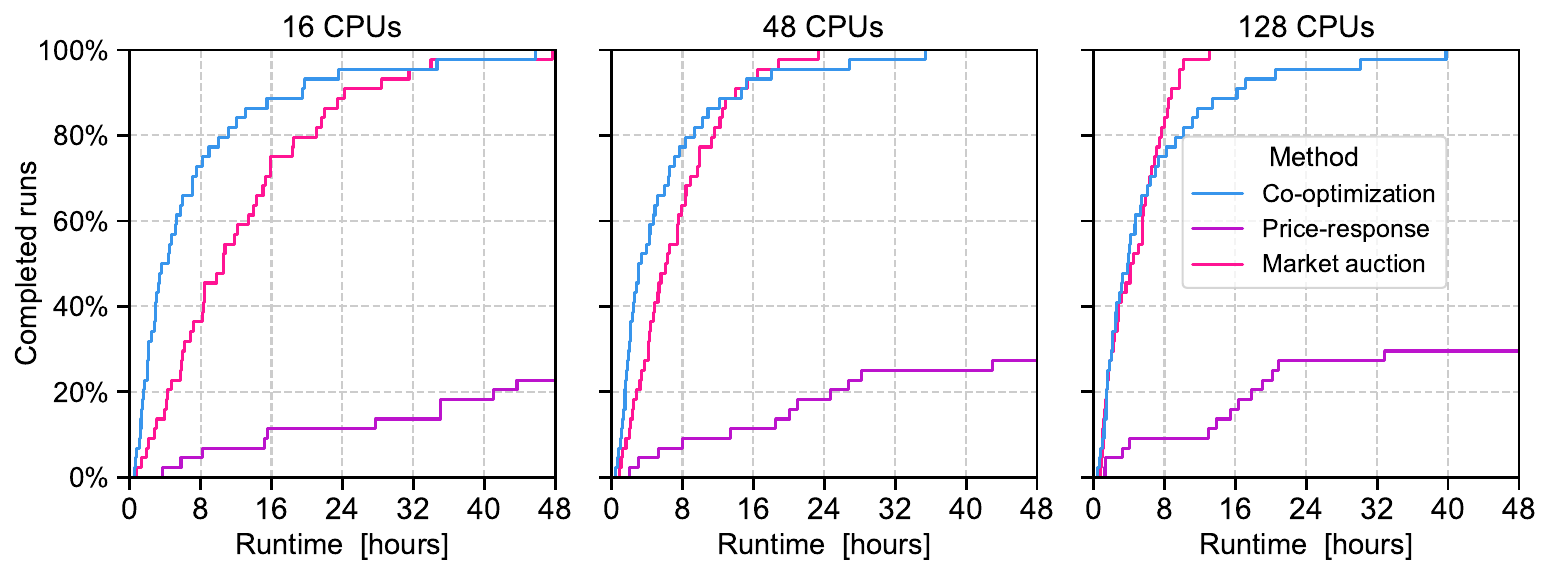}
        \label{ecdf_absolute_runtime_nonconvex}
    \end{subfigure}

    \begin{subfigure}[b]{\textwidth}
        \centering
        \caption{Normalized runtime per experiment}
        \includegraphics[height=0.88\textwidth]{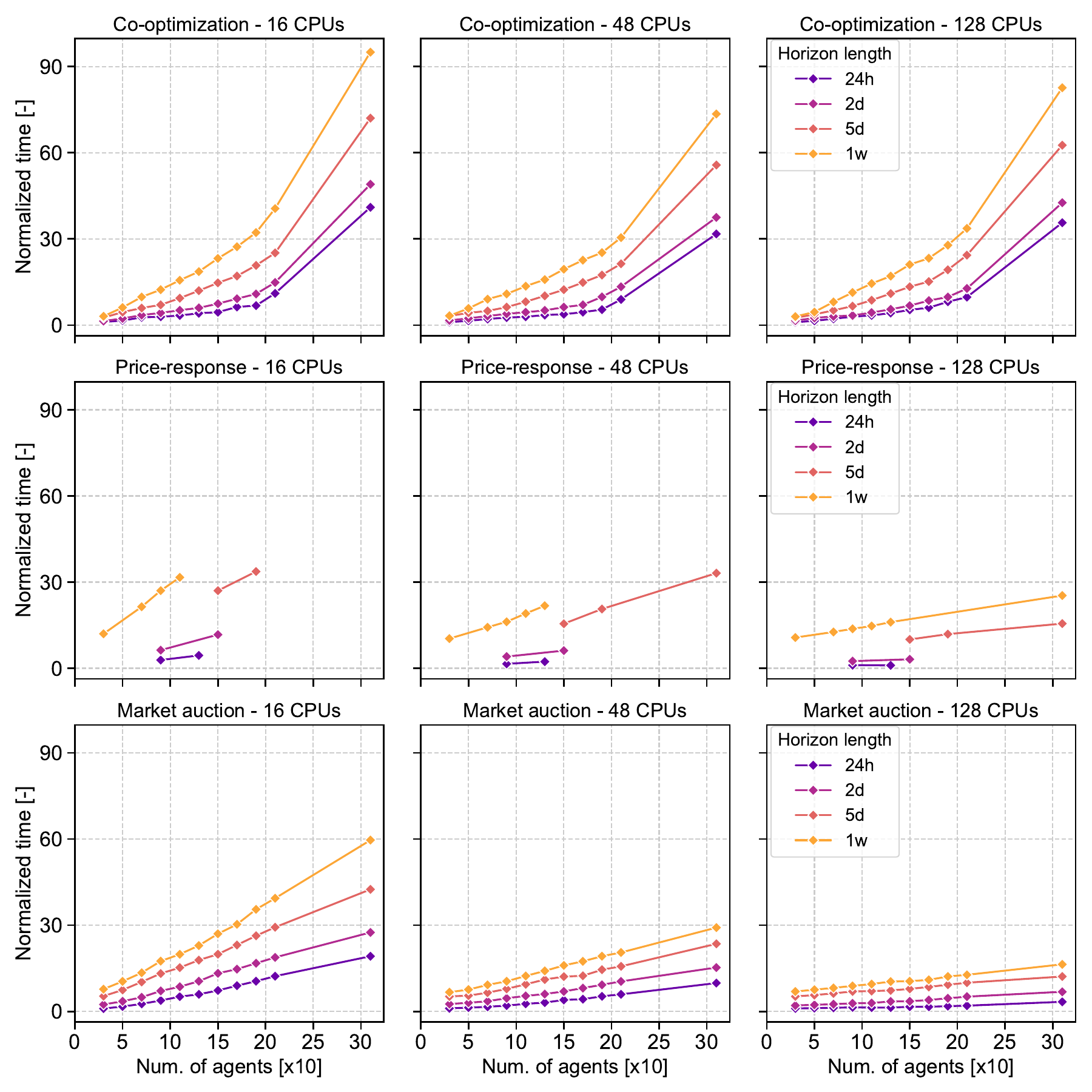}
        \label{scaling_results_nonconvex}
    \end{subfigure}

    \caption{Scaling performance under the storage charge–discharge complementarity non-convexity, across the 44 non-convex (MILP) problem instances and three CPU configurations: (a) empirical cumulative distribution of wall-clock runtime; (b) per-method minimum-time-normalized runtime. Price-response converges within the wall-clock limit on only ~27\% of instances on average; only converged runs contribute points in (b), and diverged instances are omitted.}
    \label{nonconvex_runtime_and_scaling}
\end{figure*}

\added{
\subsubsection*{Behavior under non-convexity}
The results discussed so far are for the convex (LP) setting. To analyze the scaling behavior of the three coordination approaches under non-convexities (MILP), we repeat all experiments with the storage complementarity binaries introduced in the \text{Experiment Design} section. Figure~\ref{nonconvex_runtime_and_scaling} shows the empirical findings on runtime and convergence behavior.

The cumulative distribution of absolute runtime in Figure~\ref{ecdf_absolute_runtime_nonconvex} reveals that both co-optimization and the market auction approach remain robust under non-convexities, completing all problem instances within the 48-hour wall-clock budget on all CPU configurations. Their runtime are consistently higher than in the convex case, indicating that the introduction of storage complementarity increases computational effort but does not fundamentally alter the relative efficiency ordering established earlier above for the convex cases. In particular, the market auction approach continues to reach full completion earlier than co-optimization across all CPU settings. In contrast, the price-response approach exhibits a substantial loss of robustness under non-convexity, completing only around 27\% of all instances within the same wall-clock budget. This fraction increases only marginally with additional CPU resources (from 22.7\% at 16 cores to 29.5\% at 128 cores), reflecting an inherent convergence limitation rather than a purely computational one.

The scaling behavior shown in Figure~\ref{scaling_results_nonconvex} further confirms that the relative scaling structure of co-optimization and market auction is essentially unchanged compared to the convex cases (Figure~\ref{scaling_results_convex}). This invariance arises because the per-method minimum-time normalization removes the uniform slowdown induced by the non-convexities. Consequently, both approaches preserve their scaling trend, with the market auction approach still being the fastest (the smallest slope). The price-response scaling curves, in contrast, are sparse because only the subset of instances that converge within the wall-clock budget contribute runtime observations; diverged instances are excluded since they do not yield any meaningful wall-clock runtime.

The limited completion rate of price-response can be traced to its convergence dynamics, as illustrated in Figure~\ref{primal_dual_residuals_nonconvex}. After an initial phase of rapid decay in primal and dual residuals, similar to the convex case, the algorithm enters a persistent oscillatory state in which primal and dual residuals fluctuate without approaching the convergence threshold. This behavior is driven by discrete state transitions in the binary storage variables, which induce abrupt changes in individual agent responses as the price signal evolves. While the dynamic-$\rho$ update rule mitigates imbalance when residuals become large, it cannot suppress these discontinuous flips, leading to sustained non-convergent cycles.

Overall, the non-convex results in Figure~\ref{nonconvex_runtime_and_scaling} confirm the trends observed in the convex setting while highlighting a key distinction in robustness: co-optimization and market auction remain computationally stable under non-convexities, whereas price-response loses a substantial fraction of solvable instances due to convergence failure. This reinforces the interpretation that market auction is the only coordination architecture among the three that combines scalability with robustness to non-convex operational constraints, while co-optimization remains reliable but computationally more demanding, and price-response is sensitive to discrete state dynamics in the underlying optimization problems.
}

% ====================================== non_convex_runtime_and_scaling ==========================================
% ================================================================================================================

\begin{figure*}[htbp]
    \centering
    \captionsetup{justification=centering}

    \begin{subfigure}[b]{0.49\textwidth}
        \centering
        \caption{Price-response}
        \includegraphics[width=\textwidth]{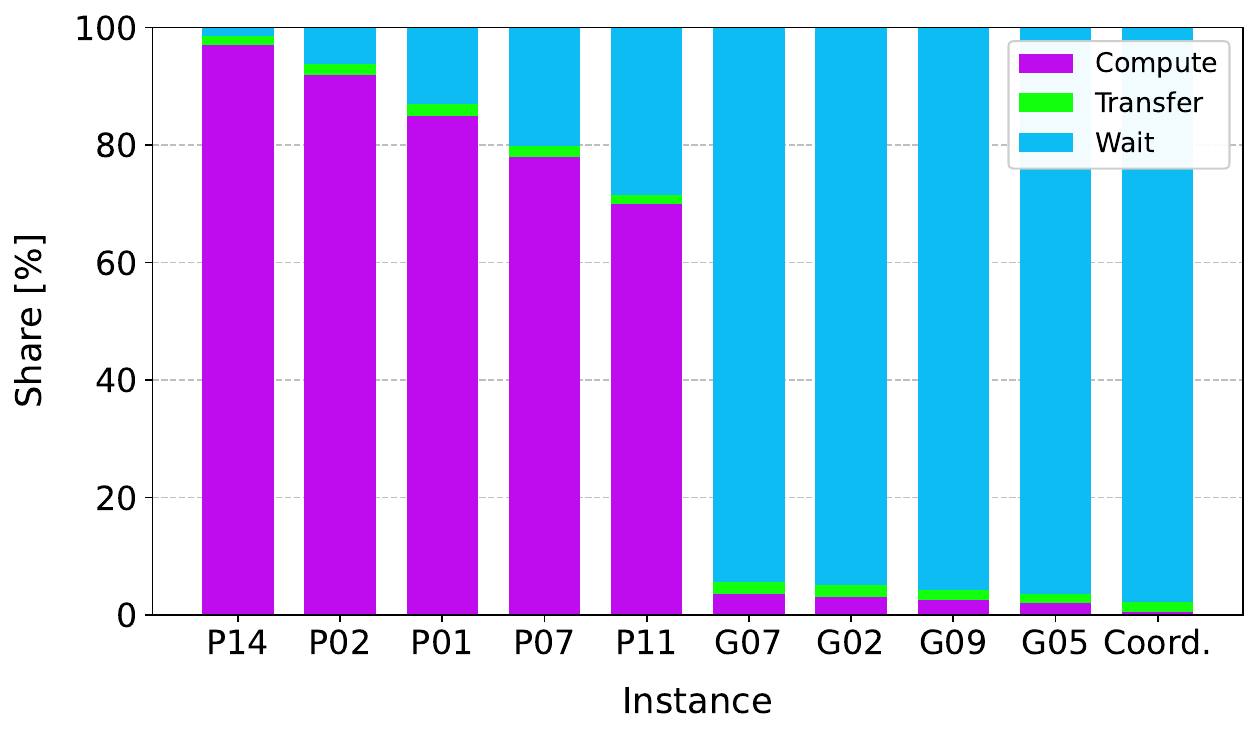}
        \label{events_per_instance_pr}
    \end{subfigure}
    \hfill
    \begin{subfigure}[b]{0.49\textwidth}
        \centering
        \caption{market auction}
        \includegraphics[width=\textwidth]{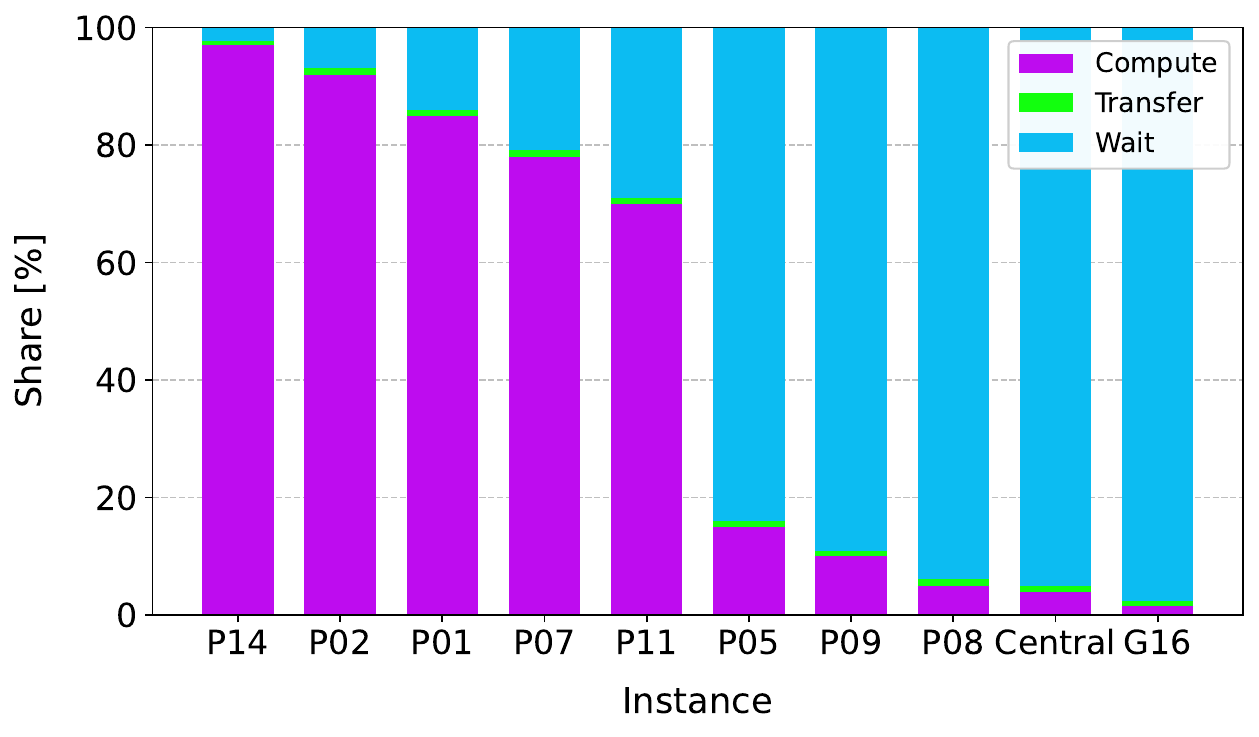}
        \label{events_per_instance_ma}
    \end{subfigure}

    \caption{Per-instance share of wall-clock time spent in compute, transfer, and wait for (a) price-response and (b) market auction averaged over all convex runs. Each plot shows the five highest-compute and five lowest-compute instances. Coord. = coordinator agent in the price-response approach; Central = central market-clearing in the market auction approach.}
    \label{events_per_instance}
\end{figure*}

\added{
\subsubsection*{Decomposition of runtime}
To give a better indication of time spent in model communication (information exchange) and actual computation, we decompose the wall-clock runtime of the two model coupling approaches into computation, communication, and waiting components. Figure~\ref{events_per_instance} reports the per-instance wall-clock breakdown for both the price-response and market auction approaches on the smallest problem instance. For both approaches, the communication share is very small (approximately 1.8\% for price-response and 1.0\% for market auction); runtime is overwhelmingly dominated by the local computations of the slowest agents and the waiting time of the fastest agents, rather than by inter-model data exchange (communication). In both model coupling approaches, the slowest agent (P14) spends more than 95\% of its runtime on local computation, while the fastest agents (the coordinator agent in price-response and G16 in market auction) spend less than 5\% on computation, remaining idle for most of the simulation while waiting for slower agents to complete.

What makes a particular submodel slow in our experiments is primarily the size of its asset portfolio and, most importantly, whether it owns storage and/or flexible demand (modeled as a store, as described in the \textit{Methodology} section). Storage assets and flexible demand introduce additional variables and constraints with strong intertemporal coupling across the optimization horizon. This translates directly into longer solve times and explains the heavy-tailed compute distribution visible in Figure~\ref{events_per_instance}: the slowest agents in both Figures~\ref{events_per_instance_pr} and~\ref{events_per_instance_ma} are prosumers that own both battery and thermal storage, whereas the fastest agents neither own storage nor solve computationally intensive scheduling problems. In the price-response approach, the coordinator only performs imbalance aggregation and price updates, while in the market auction approach, G16 submits a single bid block equal to its available capacity at marginal cost.

Therefore, the overall runtime of both model coupling approaches is ultimately constrained by the slowest submodel. This implies that further improvements in computational efficiency can effectively be achieved by targeting the computational bottlenecks of the slowest submodel, for example, through model reduction, temporal aggregation, warm-starting, decomposition of storage scheduling, or surrogate representations of computationally intensive flexibility assets.
}

% ========================================== Decomposition of runtime ===================================
% =======================================================================================================

\added{
\subsubsection*{Memory usage}
Beyond runtime, scalability also depends on memory usage, since memory exhaustion is a common failure mode for large-scale monolithic energy system optimization models. Figure~\ref{peak_memory_scaling} reports the peak memory usage (Maximum Resident Set Size, RSS) of the three coordination approaches across all convex problem instances and CPU configurations. We do not show a separate non-convex memory figure because the introduction of non-convexities (through storage-complementarity binaries) does not materially change the qualitative memory scaling behavior observed in the convex case: co-optimization remains the most memory-intensive approach, while the market auction approach exhibits the slowest memory growth.

Two trends stand out from Figure~\ref{peak_memory_scaling}. First, peak memory usage increases with problem size for all three approaches, as expected. Second, and more importantly, peak memory also increases with the number of CPU cores allocated to the solver, but the rate of increase differs substantially across the three coordination architectures. This growth is steepest for co-optimization, more moderate for the price-response approach, and smallest for the market auction approach. The mechanism behind this pattern is tied to how the solver (Gurobi's Barrier solver method)  parallelizes computations. Under the Barrier method (used in our experiments; see the \textit{Experiment Design} section), each additional thread allocates its own working memory for the sparse Cholesky factorization of the optimization problem \cite{Gurobi_Threads_Memory}. The relevant problem, however, differs substantially across the three coordination approaches. In co-optimization, all threads operate on a single large monolithic factorization, so each added thread compounds the memory requirement of the same large problem. In the price-response approach, the problem is decomposed across agents, so each thread works on a much smaller per-agent subproblem with correspondingly smaller working memory. The market auction approach decomposes the problem even further by limiting each satellite optimization to a rolling time window, thereby producing the smallest per-thread working set among the three approaches.

In other words, the deeper the decomposition of the optimization problem (across agents and across time), the smaller the per-thread memory footprint and the slower the growth in memory consumption as additional CPU cores are introduced. These results reinforce the runtime analysis above: the model coupling approaches\textemdash and especially the market auction approach\textemdash not only scale better computationally, but are also substantially less susceptible to memory exhaustion at large problem sizes.
}

% ====================================== Peak memory usage (RSS) ========================================
\added{\begin{figure}[htbp]
    \centering
    \includegraphics[width=0.9\textwidth]{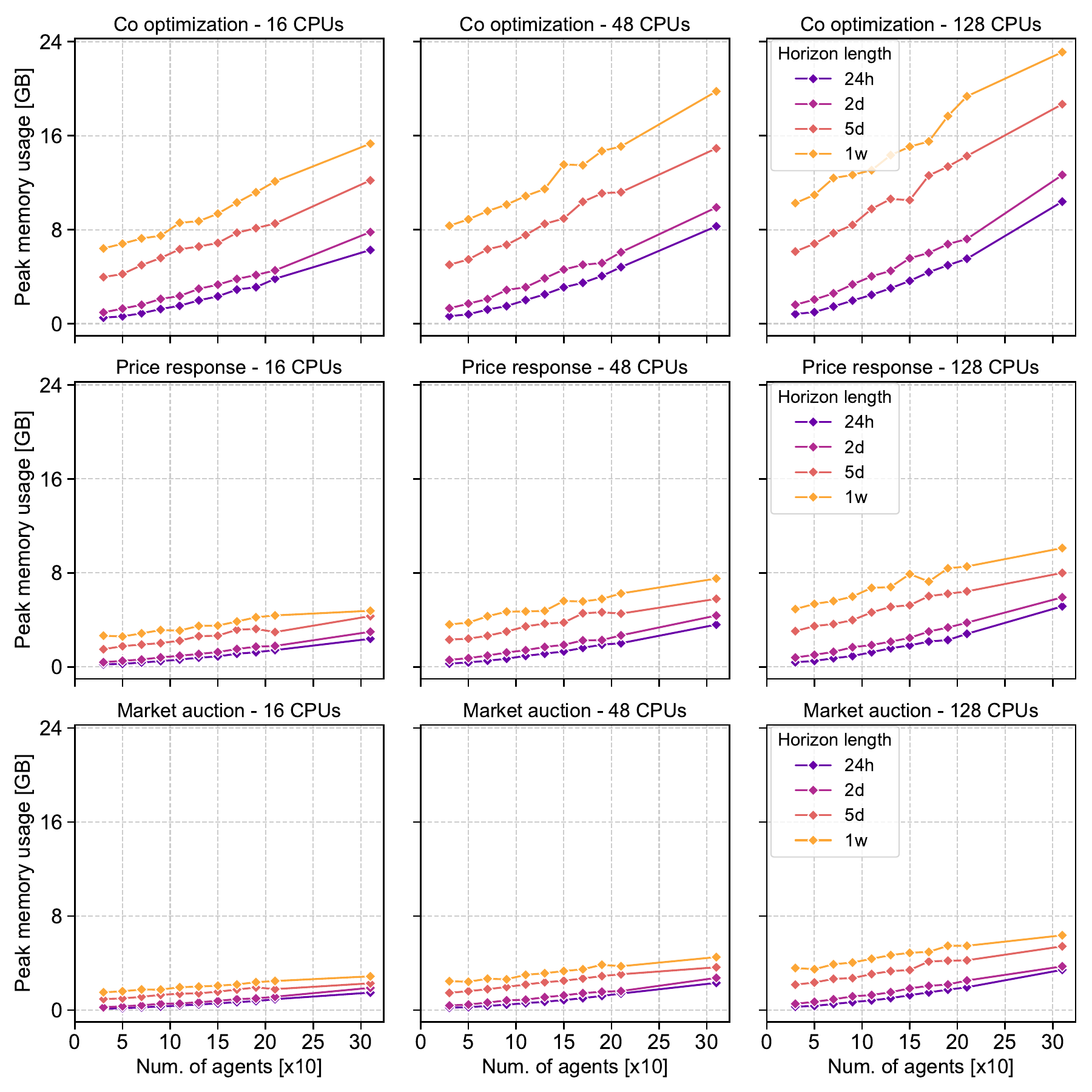}
    \vspace{2em}
    \caption{Peak memory usage (Maximum Resident Set Size, RSS) of the three coordination approaches across all 44 problem instances and three CPU configurations. Memory grows with both problem size and the number of CPU cores allocated to the solver; the growth rate with cores is steepest for co-optimization, more moderate for the price-response approach, and slowest for the market auction approach.}
    \label{peak_memory_scaling}
\end{figure}}
% =======================================================================================================

\added{%
\subsection*{Nuance in model selection}
So far, we have examined the three coordination approaches across several dimensions: optimality, autonomy, privacy, scalability in both runtime and memory, and behavior under non-convexities. Taken together, these findings do not identify a single ``best'' approach; rather, they reveal important trade-offs that make the choice of coordination architecture dependent on the modeler's objectives and constraints. We now synthesize the empirical evidence presented thus far into practical recommendations for energy system modelers. To do so, we develop a decision tree (Figure~\ref{decision_tree}) that guides the selection of a coordination approach through five sequential questions, ordered by how decisive the answer tends to be.

\begin{figure*}[htbp]
    \centering
    \includegraphics[width=0.82\textwidth]{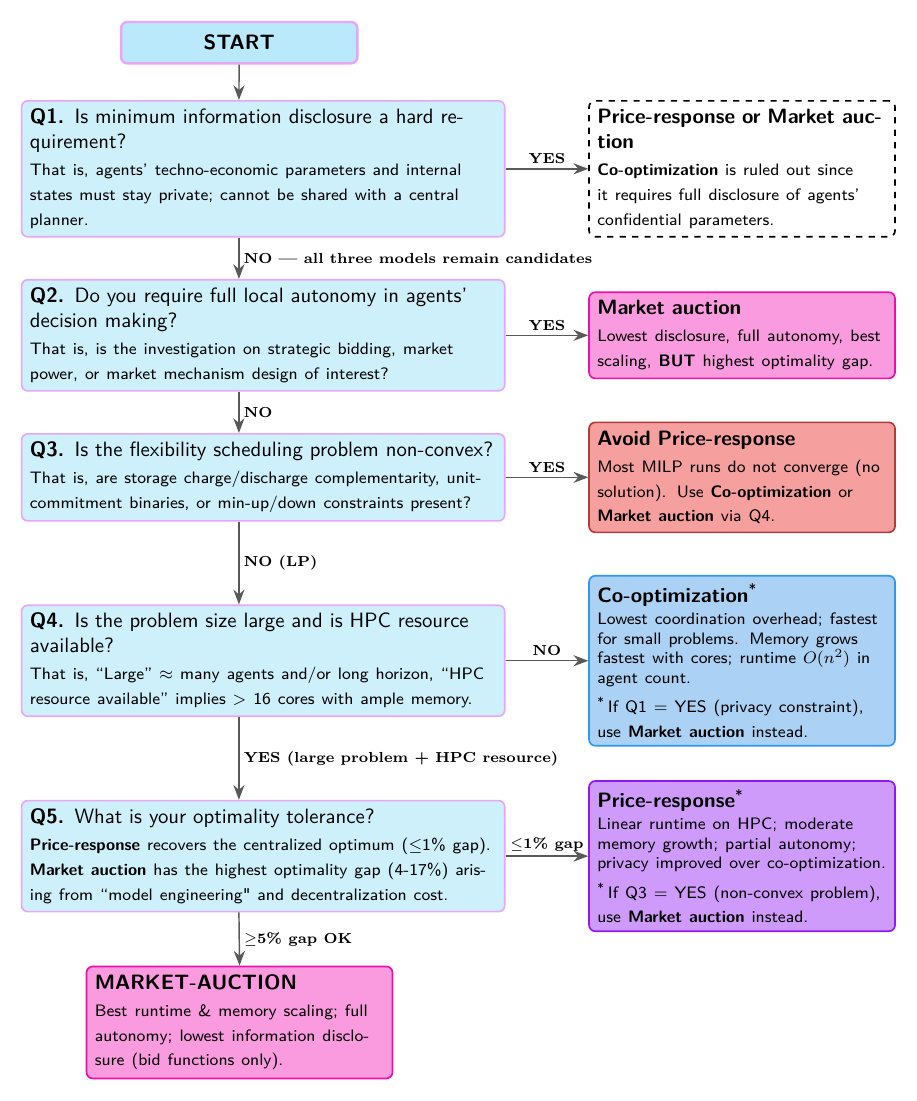}
    \vspace{2em}
    \caption{Decision tree for selecting a method for modeling flexibility scheduling in MIES. Five sequential questions route the modeler to one of three recommended flexibility modeling approaches: co-optimization, price-response, or market auction. Q1 (privacy) rules out co-optimization when minimum information disclosure is required and defers the final choice to the remaining questions; Q2 (autonomy) routes directly to market auction when full agent autonomy is required in the model; Q3 (non-convexity) flags price-response as unreliable for MILP problems and routes back through Q4; Q4 (problem size and HPC availability) selects co-optimization for small problems or modest hardware and otherwise continues to Q5; and Q5 (optimality tolerance) resolves the final choice between price-response and market auction. Asterisked leaves carry footnotes that override the default recommendation when an upstream constraint applies.}
    \label{decision_tree}
\end{figure*}

At a high level, the recommendation is as follows. If minimum information disclosure is a hard requirement (Q1)\textemdash for example, when modeling residential prosumers whose techno-economic parameters and internal states cannot be shared with a central planner\textemdash co-optimization is ruled out, and the final choice between price-response and market auction is deferred to the remaining questions below. If full local autonomy in agents' decision making is required (Q2)\textemdash for example, when studying strategic bidding, market power, or market mechanism design\textemdash market auction is the only architecture that satisfies this constraint; however, this comes at the cost of a measurable optimality gap (approximately 4--5\% in the best case, and up to 17\% in the worst case, see Figure~\ref{optimality_gap_decomposition}). When non-convexities are present in the problem definition (Q3)\textemdash for example, storage charge--discharge complementarity\textemdash the price-response approach should be avoided since only approximately 27\% of MILP instances converge within our wall-clock budget; the choice then collapses to co-optimization or market auction depending on problem size and available computing resources (Q4). When none of the above constraints applies, and the problem is convex, the choice is driven by the ratio between problem size and available computing resources: for small problems, or for larger problems on modest hardware, co-optimization remains the most efficient option since neither model coupling approach can outpace it without parallel cores (Q4); for larger problems on HPC infrastructure ($>$ 16 cores with ample memory), the choice between price-response and market auction is governed by the optimality tolerance the modeler is willing to accept (Q5)\textemdash price-response when a tight ($\leq$1\% gap) match with the centralized optimum is required, market auction when a $\geq$5\% gap is acceptable in exchange for the best runtime and memory scaling of the three approaches.

% Four caveats should be read alongside the tree: the scope bound itself; the one-time engineering cost of each approach (amortized by the open-source releases of Demoses-admm and Annular); the convergence-guarantee distinction (price-response converges only for convex problems; co-optimization and market auction always return a solution within their solver time budget); and the fact that bid-strategy and market-design experiments are uniquely possible under the market auction approach, because the bidding strategy is an explicit modeling object at the satellite layer that can be substituted without modifying the coordination architecture.

}

\added{
\subsection*{Transferability of findings to spatially-resolved systems}
The benchmark presented above adopts a copper-plate representation of the electricity system, in line with both European zonal market design and our objective of isolating the coordination layer from confounding network effects as introduced earlier in the \textit{Methodology} section. Similarly, the recommendations from Figure~\ref{decision_tree} are also based on the same copper-plate electricity system assumption. Here, we discuss the potential changes when this assumption is relaxed, the boundaries within which our quantitative claims should be interpreted, and the implications for the recommendations presented above.

Network constraints alter the coordination problem along three dimensions. First, the market-clearing price ceases to be a one-dimensional (time-indexed) vector and becomes a two-dimensional (time- and location-indexed) vector of locational marginal prices, so the information-exchange protocol of every coordination approach must be extended accordingly. For co-optimization, this is mechanical: the system-wide power balance (Equation~\ref{system_electric_power_balance_cons}) is replaced by a set of nodal balances coupled by line-flow constraints, and the cost-optimal dispatch is recovered from a larger but structurally similar problem. For the price-response approach, the consensus variable becomes a nodal time-indexed imbalance vector; the algorithm extends naturally to this setting, but at the cost of a higher-dimensional consensus problem and, in general, an expected slower convergence. For the market auction approach, the central market-clearing must be replaced by a network-constrained clearing analogous to standard nodal locational marginal pricing markets, while the satellite bid functions can, in principle, remain unchanged provided agents' bids are interpreted at their connection nodes.

Second, the feasible dispatch region of each individual agent becomes coupled to the network state reflected by the level of utilization (or congestion). In small problem instances with few large agents, this can introduce strategic incentives that the price-taking bidding heuristic in Algorithm~\ref{bid_generation_algorithm} does not capture. The qualitative finding that the market auction approach trades a measurable optimality gap for full local autonomy is, however, expected to persist; if anything, the gap is likely to grow under network constraints because locational price signals create additional opportunities for myopic dispatch decisions to diverge from the system optimum.

Third, our scalability claims should be interpreted as upper bounds on the speedup achievable in the network-constrained case. The decomposition advantage of the market auction approach\textemdash which partitions the problem across both space (agents) and time (rolling market windows)\textemdash relies on the fact that, within a clearing window, satellite problems are independent. Adding network constraints reintroduces a coupling between satellites at the central clearing step, but does not destroy the decomposition: the satellite problems themselves remain independent and parallelizable. The price-response approach is more directly affected, since its iterative consensus loop must now reach agreement on a time and location-indexed vector. Consequently, we expect the relative ranking of the three approaches by absolute runtime to be preserved on hardware with sufficient parallelism, but the magnitude of the speedup reported here (30\% on a 16-core machine and up to 70\% on a 128-core machine) is specific to the copper-plate setting and should not be extrapolated quantitatively to spatially-resolved studies.

In sum, the conceptual conclusions of this work\textemdash that model coupling provides a viable paradigm for preserving the autonomy and privacy of flexibility providers, that the price-response approach recovers the centralized optimum at the cost of iterative overhead, and that the market auction approach offers the most favorable scaling behavior at a measurable optimality cost\textemdash are expected to carry over to network-constrained settings. The specific numerical claims are bounded to the LP/MILP, copper-plate setting examined here, and we leave a quantitative assessment under network constraints to future work.
}

\section*{Conclusion}
The increasing penetration of distributed energy resources and the resulting interdependencies across energy carriers are transforming energy systems into highly complex, multi-carrier integrated energy systems (MIES). While the large-scale integration of distributed energy resources offers new sources of flexibility to cope with the variability of renewable energy, such as wind and solar, and to manage grid congestion, coordinating flexibility across many heterogeneous, autonomous prosumers remains a fundamental methodological challenge. Conventional monolithic co-optimization, while theoretically optimal, fails to adequately capture the autonomy and preserve the privacy of flexibility providers\textemdash who are increasingly participating in active demand-response\textemdash and faces significant scalability limitations due to its centralized coordination architecture.

In this work, we investigated runtime model coupling as an alternative paradigm for modeling the coordination of flexibility scheduling in MIES. We argued that model coupling can overcome the limitations of monolithic approaches by \replaced{preserving}{enhancing} the autonomy and privacy of flexibility providers\textemdash since coupled models locally optimize their decisions and only share interface variables\textemdash and also promises better scalability since coupled models can be simulated in parallel. We introduced two model coupling approaches: price-response and market auction, which can be used to couple many energy models at runtime to effectively capture the dynamic, mutually dependent interactions among flexibility providers. These two approaches mainly differ in their coordination architecture, coordination mechanism, information exchange protocol, and the extent to which autonomy is retained in agents' local decision-making. We performed a systematic quantitative benchmark of these two approaches with monolithic co-optimization under varying problem sizes, complexities, and computing infrastructures.

The results reveal significant trade-offs between optimality, autonomy, and scalability. Co-optimization can compute the theoretical system-wide optimum and performs well for small problem instances, but scales poorly as the number of agents and the temporal dimension increase. The price-response model coupling approach improves scalability while preserving a notion of coordinated optimality. However, its iterative nature introduces computational overhead that limits performance gains, particularly for large-scale problems with long time series, and faces convergence issues in the presence of non-convexities. In contrast, the market-auction approach achieves the most favorable scalability (even under non-convexities) through the decomposition of the problem in space (agents) and time, and its non-iterative coordination mechanism; however, the speedup comes at the cost of "theoretical" optimality. Nevertheless, modeling coupling offers an approach to balance realism (preservation of agents' autonomy and privacy) with computational tractability, improving \replaced{computational efficiency by 30\% under limited computational resources, and up to 70\% with additional computational resources}{cumulative batch throughput by approximately 30\% on a 16-core machine and up to 70\% on a 128-core HPC node, relative to co-optimization}. We emphasize that the reported quantitative gains are specific to the copper-plate LP/MILP benchmark considered in this study, and therefore, more work is needed to assess the extent to which these findings generalize to network-constrained formulations with nonlinear AC power flow models.

Nevertheless, the findings we uncover here advance the discussion on modeling future energy systems dominated by distributed energy resources by providing both conceptual and practical guidance, supported by empirical evidence, on how to coordinate flexibility scheduling in increasingly decentralized and integrated energy systems. By conducting a systematic quantitative benchmark of different coordination approaches, we provide empirical insights and recommendations to help energy system modelers select the coordination model that matches their modeling priorities and constraints, model scale and complexity, and available computing resources. In particular, our results suggest that co-optimization remains the preferred choice for small-scale problems and for settings in which preserving agents' autonomy and privacy is not a hard constraint; price-response is most suitable when partial autonomy in agents' decision-making, minimal information disclosure, and textbook optimality outcomes are desired, provided the problem remains convex; and market auction is preferable when full autonomy in agents' decision-making, minimal information disclosure, scalability, robustness to non-convexities, or the study of market behavior and policy design are primary modeling objectives, and near-optimality is acceptable. Beyond introducing runtime model coupling as a new paradigm for modeling flexibility in MIES, this work also helps avoid time-intensive trial-and-error exploration by offering clear, evidence-based guidance on the strengths, limitations, and trade-offs of different coordination approaches across both theoretical and practical performance metrics. Finally, to support further research and application, the proposed approaches have been formalized and implemented as open-source software tools, enabling the energy modeling community to reproduce, extend, and adapt these approaches in their own direction of interest.

% We highlight that these gains are batch-throughput gains across all problem instances rather than per-instance speed-ups; for a single large simulation run on the 128-core machine, the per-instance speed-up is more modest (a factor of approximately 2--3 for the largest problem sizes)

% \added{Finally, we restate the scope within which our quantitative claims should be interpreted. All experiments in this paper are conducted in an LP and MILP setting under a copper-plate (zonal) representation of the electricity network. The cumulative batch-throughput gains of 30--70\% and the optimality gap of approximately 10\% reported above are bounded to that setting; they may not directly translate to non-linear AC power flow models or to spatially-resolved networks (see the \emph{Scope and transferability of the findings} subsection in the Results and Discussion section for a detailed discussion of which qualitative findings are expected to carry over and which quantitative claims are not). Within the LP/MILP, copper-plate setting, however, the qualitative finding\textemdash that model coupling provides a viable architectural alternative to monolithic co-optimization for large-scale, autonomy- and privacy-aware flexibility scheduling in MIES\textemdash carries over robustly to the experimental regimes and computing infrastructures we examined.}

%\newpage

\bibliography{07_References}

@article{PONCELET_2020_AE,
title = {Unit commitment constraints in long-term planning models: Relevance, pitfalls and the role of assumptions on flexibility},
journal = {Applied Energy},
volume = {258},
pages = {113843},
year = {2020},
issn = {0306-2619},
doi = {https://doi.org/10.1016/j.apenergy.2019.113843},
author = {Kris Poncelet and Erik Delarue and William D’haeseleer}}

@article{PATTEEUW_2015_AE,
author = {Dieter Patteeuw and Kenneth Bruninx and Alessia Arteconi and Erik Delarue and William D’haeseleer and Lieve Helsen},
journal = {Applied Energy},
title = {Integrated modeling of active demand response with electric heating systems coupled to thermal energy storage systems},
volume = {151},
pages = {306-319},
year = {2015},
issn = {0306-2619},
doi = {https://doi.org/10.1016/j.apenergy.2015.04.014}}

@article{WANG_2019_AE,
title = {Operation optimization of regional integrated energy system based on the modeling of electricity-thermal-natural gas network},
journal = {Applied Energy},
volume = {251},
pages = {113410},
year = {2019},
issn = {0306-2619},
doi = {https://doi.org/10.1016/j.apenergy.2019.113410},
author = {Yongli Wang and Yudong Wang and Yujing Huang and Jiale Yang and Yuze Ma and Haiyang Yu and Ming Zeng and Fuwei Zhang and Yanfu Zhang}}

@ARTICLE{Martinez_2019_TSG,
  author={Martínez Ceseña, Eduardo Alejandro and Mancarella, Pierluigi},
  journal={IEEE Transactions on Smart Grid}, 
  title={Energy Systems Integration in Smart Districts: Robust Optimisation of Multi-Energy Flows in Integrated Electricity, Heat and Gas Networks}, 
  year={2019},
  volume={10},
  number={1},
  pages={1122-1131},
  doi={10.1109/TSG.2018.2828146}}

@ARTICLE{Mohsenian_2010_TSG,
  author={Mohsenian-Rad, Amir-Hamed and Wong, Vincent W. S. and Jatskevich, Juri and Schober, Robert and Leon-Garcia, Alberto},
  journal={IEEE Transactions on Smart Grid}, 
  title={Autonomous Demand-Side Management Based on Game-Theoretic Energy Consumption Scheduling for the Future Smart Grid}, 
  year={2010},
  volume={1},
  number={3},
  pages={320-331},
  doi={10.1109/TSG.2010.2089069}}

@article{MANCARELLA_2014_EGY,
title = {MES (multi-energy systems): An overview of concepts and evaluation models},
journal = {Energy},
volume = {65},
pages = {1-17},
year = {2014},
issn = {0360-5442},
doi = {https://doi.org/10.1016/j.energy.2013.10.041},
author = {Pierluigi Mancarella},
}

@software{doh_dinga_2025_price_response,
  author       = {Doh Dinga, Christian and
                  van Rijn, Sander},
  title        = {DEMOSES-ADMM},
  month        = apr,
  year         = 2025,
  publisher    = {Zenodo},
  version      = {v0.1.0},
  doi          = {10.5281/zenodo.15116423},
  url          = {https://doi.org/10.5281/zenodo.15116423},
}

@software{Dinga_2025_coupling,
  author       = {Doh Dinga, Christian and
                  van Rijn, Sander},
  title        = {{Annular v0.2.0}},
  month        = may,
  year         = 2025,
  publisher    = {Zenodo},
  doi          = {10.5281/zenodo.15496488},
  url          = {https://doi.org/10.5281/zenodo.15496488}
}

@misc{Dinga_2025_smart_coordination,
  author       = {Doh Dinga, Christian and
                  van Rijn, Sander},
  title        = {On the Smart Coordination of Flexibility
                   Scheduling in Multi-carrier Integrated Energy
                   Systems
                  },
  year         = 2025,
  publisher    = {Zenodo},
  doi          = {10.5281/zenodo.15496935},
  url          = {https://doi.org/10.5281/zenodo.15496935},
}

@article{OCONNELL_2014_RSER,
title = {Benefits and challenges of electrical demand response: A critical review},
journal = {Renewable and Sustainable Energy Reviews},
volume = {39},
pages = {686-699},
year = {2014},
issn = {1364-0321},
doi = {https://doi.org/10.1016/j.rser.2014.07.098},
author = {Niamh O'Connell and Pierre Pinson and Henrik Madsen and Mark O'Malley},
}

@INPROCEEDINGS{Dinga_ISGT_2024,
  author={Dinga, Christian Doh and Van Rijn, Sander and De Vries, Laurens and Cvetkovic, Milos},
  booktitle={2024 IEEE PES Innovative Smart Grid Technologies Europe (ISGT EUROPE)}, 
  title={Coordinated flexibility scheduling in multi-carrier integrated energy systems: a model coupling approach}, 
  year={2024},
  volume={},
  number={},
  pages={1-5},
  doi={10.1109/ISGTEUROPE62998.2024.10863244}}

@ARTICLE{Bruninx_2018_TSE,
  author={Bruninx, Kenneth and Dvorkin, Yury and Delarue, Erik and D’haeseleer, William and Kirschen, Daniel S.},
  journal={IEEE Transactions on Sustainable Energy}, 
  title={Valuing Demand Response Controllability via Chance Constrained Programming}, 
  year={2018},
  volume={9},
  number={1},
  pages={178-187},
  doi={10.1109/TSTE.2017.2718735}}

@article{HAQUE_SEGN_2017,
title = {Demand response for real-time congestion management incorporating dynamic thermal overloading cost},
journal = {Sustainable Energy, Grids and Networks},
volume = {10},
pages = {65-74},
year = {2017},
issn = {2352-4677},
doi = {https://doi.org/10.1016/j.segan.2017.03.002},
author = {A.N.M.M. Haque and P.H. Nguyen and F.W. Bliek and J.G. Slootweg},
}

@ARTICLE{Li_2022_TIA,
  author={Li, Yang and Wang, Bin and Yang, Zhen and Li, Jiazheng and Li, Guoqing},
  journal={IEEE Transactions on Industry Applications}, 
  title={Optimal Scheduling of Integrated Demand Response-Enabled Community-Integrated Energy Systems in Uncertain Environments}, 
  year={2022},
  volume={58},
  number={2},
  pages={2640-2651},
  doi={10.1109/TIA.2021.3106573}}

@ARTICLE{Zhang_2019_TSG,
  author={Zhang, Xi and Strbac, Goran and Shah, Nilay and Teng, Fei and Pudjianto, Danny},
  journal={IEEE Transactions on Smart Grid}, 
  title={Whole-System Assessment of the Benefits of Integrated Electricity and Heat System}, 
  year={2019},
  volume={10},
  number={1},
  pages={1132-1145},
  doi={10.1109/TSG.2018.2871559}}

@ARTICLE{Zhang_2016_TPS,
  author={Zhang, Xiaping and Shahidehpour, Mohammad and Alabdulwahab, Ahmed and Abusorrah, Abdullah},
  journal={IEEE Transactions on Power Systems}, 
  title={Hourly Electricity Demand Response in the Stochastic Day-Ahead Scheduling of Coordinated Electricity and Natural Gas Networks}, 
  year={2016},
  volume={31},
  number={1},
  pages={592-601},
  doi={10.1109/TPWRS.2015.2390632}}

@ARTICLE{Gottwalt_2017_TSG,
  author={Gottwalt, Sebastian and Gärttner, Johannes and Schmeck, Hartmut and Weinhardt, Christof},
  journal={IEEE Transactions on Smart Grid}, 
  title={Modeling and Valuation of Residential Demand Flexibility for Renewable Energy Integration}, 
  year={2017},
  volume={8},
  number={6},
  pages={2565-2574},
  doi={10.1109/TSG.2016.2529424}}

@article{DONG_2023_AE,
title = {Refined modeling and co-optimization of electric-hydrogen-thermal-gas integrated energy system with hybrid energy storage},
journal = {Applied Energy},
volume = {351},
pages = {121834},
year = {2023},
issn = {0306-2619},
doi = {https://doi.org/10.1016/j.apenergy.2023.121834},
author = {Haoxin Dong and Zijing Shan and Jianli Zhou and Chuanbo Xu and Wenjun Chen}}

@ARTICLE{Palensky_2017_IEM,
  author={Palensky, Peter and Van Der Meer, Arjen A. and Lopez, Claudio David and Joseph, Arun and Pan, Kaikai},
  journal={IEEE Industrial Electronics Magazine}, 
  title={Cosimulation of Intelligent Power Systems: Fundamentals, Software Architecture, Numerics, and Coupling}, 
  year={2017},
  volume={11},
  number={1},
  pages={34-50},
  doi={10.1109/MIE.2016.2639825}}

@ARTICLE{Palensky_2024_PEM,
  author={Palensky, Peter and Mancarella, Pierluigi and Hardy, Trevor and Cvetkovic, Milos},
  journal={IEEE Power and Energy Magazine}, 
  title={Cosimulating Integrated Energy Systems With Heterogeneous Digital Twins: Matching a Connected World}, 
  year={2024},
  volume={22},
  number={1},
  pages={52-60},
  doi={10.1109/MPE.2023.3324886}}

@ARTICLE{Hoschle_2018_TPS,
  author={Höschle, Hanspeter and Le Cadre, Hélène and Smeers, Yves and Papavasiliou, Anthony and Belmans, Ronnie},
  journal={IEEE Transactions on Power Systems}, 
  title={An ADMM-Based Method for Computing Risk-Averse Equilibrium in Capacity Markets}, 
  year={2018},
  volume={33},
  number={5},
  pages={4819-4830},
  doi={10.1109/TPWRS.2018.2807738}}

@BOOK{Boyd_2011,
  author={Boyd, Stephen and Parikh, Neal and Chu, Eric and Peleato, Borja and Eckstein, Jonathan},
  title={Distributed Optimization and Statistical Learning via the Alternating Direction Method of Multipliers},
  Publisher={Now Foundations and Trends},
  year={2011},
  volume={},
  number={},
  pages={},
  doi={10.1561/2200000016}}

@article{Brown_2018_ENGY,
   author = {T. Brown and D. Schlachtberger and A. Kies and S. Schramm and M. Greiner},
   doi = {10.1016/j.energy.2018.06.222},
   issn = {03605442},
   journal = {Energy},
   month = {10},
   pages = {720-739},
   publisher = {Elsevier Ltd},
   title = {Synergies of sector coupling and transmission reinforcement in a cost-optimised, highly renewable European energy system},
   volume = {160},
   year = {2018},
}

@article{NEUMANN_2023_Joule,
    title = {The potential role of a hydrogen network in Europe},
    journal = {Joule},
    volume = {7},
    number = {8},
    pages = {1793-1817},
    year = {2023},
    issn = {2542-4351},
    doi = {https://doi.org/10.1016/j.joule.2023.06.016},
    author = {Fabian Neumann and Elisabeth Zeyen and Marta Victoria and Tom Brown},
}

@misc{SURF_Snellius,
  author       = {{SURF}},
  title        = {Snellius, the national supercomputer},
  url          = {https://www.surf.nl/en/services/compute/snellius-the-national-supercomputer},
}

@ARTICLE{Harder_2023_EI,
  author={Harder, Nick and Weidlich, Anke and Staudt, Philipp},
  journal={Energy Informatics}, 
  title={Finding individual strategies for storage units in electricity market models using deep reinforcement learning}, 
  year={2023},
  volume={6},
  number={},
  issn = {2520-8942},
  pages={488-502},
  doi={10.1186/s42162-023-00293-0}
}

@article{lund_2023_model_coupling_energy,
title = {Perspectives on purpose-driven coupling of energy system models},
journal = {Energy},
volume = {265},
pages = {126335},
year = {2023},
issn = {0360-5442},
doi = {https://doi.org/10.1016/j.energy.2022.126335},
author = {Miguel Chang and Henrik Lund and Jakob Zinck Thellufsen and Poul Alberg Østergaard},
}

@article{WIDL2022_segn,
    title = {Expert survey and classification of tools for modeling and simulating hybrid energy networks},
    journal = {Sustainable Energy, Grids and Networks},
    volume = {32},
    pages = {100913},
    year = {2022},
    issn = {2352-4677},
    doi = {https://doi.org/10.1016/j.segan.2022.100913},
    author = {Edmund Widl and Dennis Cronbach and Peter Sorknæs and Jaume Fitó and Daniel Muschick and Maurizio Repetto and Julien Ramousse and Anton Ianakiev},
}

@article{Frigg_2022_ae,
title = {Frigg: Soft-linking energy system and demand response models},
journal = {Applied Energy},
volume = {317},
pages = {119074},
year = {2022},
issn = {0306-2619},
doi = {https://doi.org/10.1016/j.apenergy.2022.119074},
author = {Amos Schledorn and Rune Grønborg Junker and Daniela Guericke and Henrik Madsen and Dominik Franjo Dominković},
}

@article{ZHANG_2024_soft_linking,
title = {An novel soft-linking model: Coordinating the regulative flexibility and operational stability of pumped storage units},
journal = {Applied Energy},
volume = {374},
pages = {123942},
year = {2024},
issn = {0306-2619},
doi = {https://doi.org/10.1016/j.apenergy.2024.123942},
author = {Meng Zhang and Yu Xiao and Yuanqiang Gao and Daiyu Li and Ziwen Zhao and Jianling Li and Liuwei Lei and Zhengguang Liu and Mengjiao He and Diyi Chen},
}

@article{RAJAKOVIC_2025_ecmx,
title = {Coupling of the planning and power system analysis software for high-level penetration of renewable energy sources},
journal = {Energy Conversion and Management: X},
volume = {28},
pages = {101298},
year = {2025},
issn = {2590-1745},
doi = {https://doi.org/10.1016/j.ecmx.2025.101298},
author = {Nikola Rajakovic and Bojan Ivanović and Ilija Batas Bjelic and Tomislav Rajić},
}

@article{ROSENDAL_2025_soft_linking,
title = {The benefits and challenges of soft-linking investment and operational energy system models},
journal = {Applied Energy},
volume = {385},
pages = {125512},
year = {2025},
issn = {0306-2619},
doi = {https://doi.org/10.1016/j.apenergy.2025.125512},
author = {M. Rosendal and J. Janin and T. Heggarty and D. Pisinger and R. Bramstoft and M. Münster},
}

@article{garg_2025_nsr,
  title     = {Optimal energy management of multi-carrier energy system considering uncertainty in renewable generation},
  author    = {Garg, Ankit and Niazi, K. R. and Tiwari, Shubham and Sharma, Sachin and Rawat, Tanuj},
  journal   = {Scientific Reports},
  volume    = {15},
  number    = {1},
  pages     = {25936},
  year      = {2025},
  doi       = {10.1038/s41598-025-10404-4}
}

@article{LUND_2015_review_flex,
title = {Review of energy system flexibility measures to enable high levels of variable renewable electricity},
journal = {Renewable and Sustainable Energy Reviews},
volume = {45},
pages = {785-807},
year = {2015},
issn = {1364-0321},
doi = {https://doi.org/10.1016/j.rser.2015.01.057},
author = {Peter D. Lund and Juuso Lindgren and Jani Mikkola and Jyri Salpakari},
}

@misc{irena_2018_flexibility,
  author       = {{International Renewable Energy Agency}},
  title        = {Power system flexibility for the energy transition, part 1: overview for policy makers},
  year         = {2018},
  publisher    = {International Renewable Energy Agency},
  address      = {Abu Dhabi},
  note         = {\url{https://www.irena.org/-/media/Files/IRENA/Agency/Publication/2018/Nov/IRENA_Power_system_flexibility_1_2018.pdf}}
}

@article{Strbac_2020_IOP,
doi = {10.1088/2516-1083/abb216},
year = {2020},
month = {sep},
publisher = {IOP Publishing},
volume = {2},
number = {4},
pages = {042001},
author = {Strbac, Goran and Pudjianto, Danny and Aunedi, Marko and Djapic, Predrag and Teng, Fei and Zhang, Xi and Ameli, Hossein and Moreira, Roberto and Brandon, Nigel},
title = {Role and value of flexibility in facilitating cost-effective energy system decarbonisation},
journal = {Progress in Energy},
}

@article{GHATIKAR_2016_autonomous,
title = {Distributed energy systems integration and demand optimization for autonomous operations and electric grid transactions},
journal = {Applied Energy},
volume = {167},
pages = {432-448},
year = {2016},
issn = {0306-2619},
doi = {https://doi.org/10.1016/j.apenergy.2015.10.117},
author = {Girish Ghatikar and Salman Mashayekh and Michael Stadler and Rongxin Yin and Zhenhua Liu},
}

@article{STAWSKA_2021_EP,
title = {Demand response: For congestion management or for grid balancing?},
journal = {Energy Policy},
volume = {148},
pages = {111920},
year = {2021},
issn = {0301-4215},
doi = {https://doi.org/10.1016/j.enpol.2020.111920},
author = {Anna Stawska and Natalia Romero and Mathijs {de Weerdt} and Remco Verzijlbergh},
}

@ARTICLE{Martinez_2020,
  author={Martínez Ceseña, Eduardo Alejandro and Loukarakis, Emmanouil and Good, Nicholas and Mancarella, Pierluigi},
  journal={Proceedings of the IEEE}, 
  title={Integrated Electricity– Heat–Gas Systems: Techno–Economic Modeling, Optimization, and Application to Multienergy Districts}, 
  year={2020},
  volume={108},
  number={9},
  pages={1392-1410},
  doi={10.1109/JPROC.2020.2989382}
  }

@article{li_2025_optimization,
  title     = {An optimization method for integrated demand response strategies for electricity and heat considering the uncertainty of user-side loads},
  author    = {Li, Jiaqi and Zhang, Delong and Wei, Yuheng and Zhou, Xuesong and Kong, Xiangyu and Huo, Xianxu and Pang, Chao},
  journal   = {Scientific Reports},
  volume    = {16},
  number    = {1},
  pages     = {582},
  year      = {2025},
  doi       = {10.1038/s41598-025-30090-6}
}

@article{ZHANG_2024_SETA,
title = {Optimal scheduling of electricity-gas-heat integrated energy system with coordination of flexibility and reliability},
journal = {Sustainable Energy Technologies and Assessments},
volume = {71},
pages = {103968},
year = {2024},
issn = {2213-1388},
doi = {https://doi.org/10.1016/j.seta.2024.103968},
author = {Yumin Zhang and Jingrui Li and Xingquan Ji and Ming Yang and Pingfeng Ye},
}

@article{Aswani_2018_OR,
    author = {Aswani, A. and Shen, Z. M. and Siddiq, A.},
    title = {Inverse Optimization with Noisy Data},
    journal = {Operations Research},
    year = {2018},
    volume = {66},
    issue = {3},
    pages = {870-892},
    doi = {10.1287/opre.2017.1705} 
}

@article{Borenstein_2002_AER,
  author = {Borenstein, S. and Bushnell, J. and Wolak, F. A.},
  title = {Measuring Market Inefficiencies in California's Restructured Wholesale Electricity Market},
  journal = {American Economic Review},
  year = {2002},
  volume = {92},
  issue = {5},
  pages = {1376-1405},
  doi = {10.1257/000282802762024557}
}

@article{Bruninx_2020_EE,
  author = {Bruninx, K. and Ovaere, M. and Delarue, E.},
  title = {The long-term impact of the market stability reserve on the EU emission trading system},
  journal = {Energy Economics},
  year = {2020},
  volume = {89},
  pages = {104746},
  doi = {10.1016/j.eneco.2020.104746}
}

@ARTICLE{Cramton_2017_OREP,
    author = {Cramton, Peter},
    title = {Electricity market design},
    year = {2017},
    journal = {Oxford Review of Economic Policy},
    volume = {33},
    number = {4},
    pages = {589-612},
    url = {https://econpapers.repec.org/article/oupoxford/v_3a33_3ay_3a2017_3ai_3a4_3ap_3a589-612..htm},
}

@ARTICLE{Dvorkin_2020_TPS,
  author={Dvorkin, Vladimir and Fioretto, Ferdinando and Van Hentenryck, Pascal and Pinson, Pierre and Kazempour, Jalal},
  journal={IEEE Transactions on Power Systems}, 
  title={Differentially Private Optimal Power Flow for Distribution Grids}, 
  year={2021},
  volume={36},
  number={3},
  pages={2186-2196},
  doi={10.1109/TPWRS.2020.3031314}
}

@ARTICLE{Han_2017_TAC,
  author={Han, Shuo and Topcu, Ufuk and Pappas, George J.},
  journal={IEEE Transactions on Automatic Control}, 
  title={Differentially Private Distributed Constrained Optimization}, 
  year={2017},
  volume={62},
  number={1},
  pages={50-64},
  doi={10.1109/TAC.2016.2541298}
}

@article{Vazquez_Canteli_2019_AE,
    author = {José R. Vázquez-Canteli and Zoltán Nagy},
    title = {Reinforcement learning for demand response: A review of algorithms and modeling techniques},
    journal = {Applied Energy},
    volume = {235},
    pages = {1072-1089},
    year = {2019},
    issn = {0306-2619},
    doi = {https://doi.org/10.1016/j.apenergy.2018.11.002},
}

@ARTICLE{Saad_2012_SPM,
  author={Saad, Walid and Han, Zhu and Poor, H. Vincent and Basar, Tamer},
  journal={IEEE Signal Processing Magazine}, 
  title={Game-Theoretic Methods for the Smart Grid: An Overview of Microgrid Systems, Demand-Side Management, and Smart Grid Communications}, 
  year={2012},
  volume={29},
  number={5},
  pages={86-105},
  doi={10.1109/MSP.2012.2186410}
  }

@article{Deane_2012_EGY,
    title = {Soft-linking of a power systems model to an energy systems model},
    journal = {Energy},
    volume = {42},
    number = {1},
    pages = {303-312},
    year = {2012},
    note = {8th World Energy System Conference, WESC 2010},
    issn = {0360-5442},
    doi = {https://doi.org/10.1016/j.energy.2012.03.052},
    author = {J.P. Deane and Alessandro Chiodi and Maurizio Gargiulo and Brian P. {Ó Gallachóir}},
}

@article{Gotske_2024_NComms,
    author = {Ebbe Kyhl Gøtske and Gorm Bruun Andresen and Fabian Neumann and Marta Victoria},
    title = {Designing a sector-coupled European energy system robust to 60 years of historical weather data},
    journal = {Nature Communications},
    volume = {15},
    number = {1},
    pages = {10680},
    year = {2024},
    issn = {2041-1723},
    doi = {https://doi.org/10.1038/s41467-024-54853-3},
}

@article{Glaum_2026_Joule,
    author = {Philipp Glaum and Fabian Neumann and Markus Millinger and Tom Brown},
    title = {A minimal methanol backstop for high-electrification scenarios},
    journal = {Joule},
    year = {2026},
    issn = {2542-4785},
    publisher = {Elsevier},
    doi = {https://doi.org/10.1016/j.joule.2026.102464},
}

@article{Xiong_2025_ERL,
doi = {10.1088/1748-9326/ae3846},
year = {2026},
month = {feb},
publisher = {IOP Publishing},
volume = {21},
number = {3},
pages = {034011},
author = {Xiong, Bobby and Brown, Tom and Riepin, Iegor},
title = {The role of projects of common interest in reaching Europe’s energy policy targets},
journal = {Environmental Research Letters},
}

@article{Neumann_2025_NComms,
    author = {Fabian Neumann and Johannes Hampp and Tom Brown},
    title = {Green energy and steel imports reduce Europe’s net-zero infrastructure needs},
    journal = {Nature Communications},
    volume = {16},
    number = {1},
    pages = {5302},
    year = {2025},
    issn = {2041-1723},
    doi = {https://doi.org/10.1038/s41467-025-60652-1},
}

@phdthesis{gusain_2022_dissertation,
  author       = {Gusain, D.},
  title        = {Flexibility in Multi Energy Systems: Models and Metrics to Assess Future Energy Systems},
  school       = {Delft University of Technology},
  year         = {2022},
  doi          = {10.4233/uuid:3e857b6c-b028-4f40-9625-54ca79477be8},
  type         = {PhD dissertation}
}

@ARTICLE{Kargarian_2018_distributed_vs_decentralized,
  author={Kargarian, Amin and Mohammadi, Javad and Guo, Junyao and Chakrabarti, Sambuddha and Barati, Masoud and Hug, Gabriela and Kar, Soummya and Baldick, Ross},
  journal={IEEE Transactions on Smart Grid}, 
  title={Toward Distributed/Decentralized DC Optimal Power Flow Implementation in Future Electric Power Systems}, 
  year={2018},
  volume={9},
  number={4},
  pages={2574-2594},
  doi={10.1109/TSG.2016.2614904}
}

@article{Poplavskaya_2020_AE,
  title   = {Integration of day-ahead market and redispatch to increase cross-border exchanges in the {E}uropean electricity market},
  author  = {Poplavskaya, Ksenia and Totschnig, Gerhard and Leimgruber, Fabian and Doorman, Gerard and Etienne, Gilles and de Vries, Laurens},
  journal = {Applied Energy},
  volume  = {278},
  pages   = {115669},
  year    = {2020},
  doi     = {10.1016/j.apenergy.2020.115669},
  publisher = {Elsevier}
}

@misc{Gurobi_Threads_Memory,
author       = {{Gurobi Optimization, LLC}},
title        = {Managing Threads and Memory Usage in Gurobi},
year         = {2025},
howpublished = {\url{https://support.gurobi.com/hc/en-us/articles/42058344459409}},
note         = {Accessed: 2026-05-29}
}

@software{DohDinga_2025_cronian,
author = {Doh Dinga, Christian and van Rijn, Sander},
title = {cronian},
year = {2025},
version = {0.3.2},
publisher = {Zenodo},
doi = {10.5281/zenodo.13142375},
url = {https://doi.org/10.5281/zenodo.13142375}
}

@Article{mosaik_2019,
AUTHOR = {Steinbrink, Cornelius and Blank-Babazadeh, Marita and El-Ama, André and Holly, Stefanie and Lüers, Bengt and Nebel-Wenner, Marvin and Ramírez Acosta, Rebeca P. and Raub, Thomas and Schwarz, Jan Sören and Stark, Sanja and Nieße, Astrid and Lehnhoff, Sebastian},
TITLE = {CPES Testing with mosaik: Co-Simulation Planning, Execution and Analysis},
JOURNAL = {Applied Sciences},
VOLUME = {9},
YEAR = {2019},
NUMBER = {5},
ARTICLE-NUMBER = {923},
URL = {https://www.mdpi.com/2076-3417/9/5/923},
ISSN = {2076-3417},
DOI = {10.3390/app9050923}
}

@ARTICLE{HELICS_2024,
author={Hardy, Trevor D. and Palmintier, Bryan and Top, Philip L. and Krishnamurthy, Dheepak and Fuller, Jason C.},
journal={IEEE Access}, 
title={HELICS: A Co-Simulation Framework for Scalable Multi-Domain Modeling and Analysis}, 
year={2024},
volume={12},
number={},
pages={24325-24347},
doi={10.1109/ACCESS.2024.3363615}
}

@article{GUSAIN_EnergySym,
title = {Simplifying multi-energy system co-simulations using energysim},
journal = {SoftwareX},
volume = {18},
pages = {101021},
year = {2022},
issn = {2352-7110},
doi = {https://doi.org/10.1016/j.softx.2022.101021},
url = {https://www.sciencedirect.com/science/article/pii/S2352711022000255},
author = {Digvijay Gusain and Milos̆ Cvetković and Peter Palensky},
}

% \newpage
\bigskip
\bigskip

\section*{Acknowledgments}
This publication is part of the DEMOSES project, financed by the Dutch Research Council (NWO) under grant ID ESI.2019.004. This work used the Dutch national e-infrastructure with the support of the SURF Cooperative, using grant no. EINF-11996.

\section*{Author contributions statement}
C.D.D. and M.C; conceptualization, C.D.D. and S.V.R; methodology and software, C.D.D.; data curation, investigation and visualization, C.D.D.; writing-–original draft, C.D.D. and and S.V.R; writing-–review \& editing, L.J.D.V. and M.C.; validation, L.J.D.V. and M.C.; funding acquisition, L.J.D.V. and M.C.; resources, L.J.D.V. and M.C.; supervision.

\section*{Declarations}

\section*{Competing interests}
The authors declare no competing interests.

\section*{Data and code availability}
The code to reproduce the experiment results and visualizations in this study is available on GitLab: 
\url{https://gitlab.tudelft.nl/demoses/demoses-scaling-benchmarks}, 
and is archived on Zenodo: 
\url{https://zenodo.org/records/15496935}.

\section*{Additional information}
Correspondence and requests for further information, resources, and materials should be directed to and will be fulfilled by Christian Doh Dinga (c.dohdinga@tu.delft.nl).

\end{document}